\documentclass[modern]{aastex63}

\usepackage{textcomp}

\accepted{September 25, 2020}
\submitjournal{Planetary Science Journal}

\shorttitle{Analysis of a Ryugu rock}

\begin{document}

\title{Spectrophotometric analysis of the Ryugu rock seen by {\sc mascot}: Searching for a carbonaceous chondrite analog}

\correspondingauthor{Stefan Schr\"oder}
\email{stefanus.schroeder@dlr.de}

\author[0000-0003-0323-8324]{Stefan Schr\"oder}
\affiliation{Deutsches Zentrum f\"ur Luft- und Raumfahrt (DLR) \\
Institute of Planetary Research \\
12489 Berlin, Germany}

\author[0000-0002-0675-1177]{Katharina Otto}
\affiliation{Deutsches Zentrum f\"ur Luft- und Raumfahrt (DLR) \\
Institute of Planetary Research \\
12489 Berlin, Germany}

\author{Hannah Scharf}
\affiliation{Deutsches Zentrum f\"ur Luft- und Raumfahrt (DLR) \\
Institute of Planetary Research \\
12489 Berlin, Germany}

\author[0000-0002-4148-1926]{Klaus-Dieter Matz}
\affiliation{Deutsches Zentrum f\"ur Luft- und Raumfahrt (DLR) \\
Institute of Planetary Research \\
12489 Berlin, Germany}

\author{Nicole Schmitz}
\affiliation{Deutsches Zentrum f\"ur Luft- und Raumfahrt (DLR) \\
Institute of Planetary Research \\
12489 Berlin, Germany}

\author[0000-0002-1430-9223]{Frank Scholten}
\affiliation{Deutsches Zentrum f\"ur Luft- und Raumfahrt (DLR) \\
Institute of Planetary Research \\
12489 Berlin, Germany}

\author[0000-0002-0457-3872]{Stefano Mottola}
\affiliation{Deutsches Zentrum f\"ur Luft- und Raumfahrt (DLR) \\
Institute of Planetary Research \\
12489 Berlin, Germany}

\author{Frank Trauthan}
\affiliation{Deutsches Zentrum f\"ur Luft- und Raumfahrt (DLR) \\
Institute of Planetary Research \\
12489 Berlin, Germany}

\author[0000-0001-8231-1109]{Alexander Koncz}
\affiliation{Deutsches Zentrum f\"ur Luft- und Raumfahrt (DLR) \\
Institute of Planetary Research \\
12489 Berlin, Germany}

\author{Harald Michaelis}
\affiliation{Deutsches Zentrum f\"ur Luft- und Raumfahrt (DLR) \\
Institute of Planetary Research \\
12489 Berlin, Germany}

\author[0000-0002-9759-6597]{Ralf Jaumann}
\affiliation{Deutsches Zentrum f\"ur Luft- und Raumfahrt (DLR) \\
Institute of Planetary Research \\
12489 Berlin, Germany}

\author[0000-0002-9759-6597]{Tra-Mi Ho}
\affiliation{Deutsches Zentrum f\"ur Luft- und Raumfahrt (DLR) \\
Institute of Space Systems \\
28359 Bremen, Germany}

\author[0000-0002-4625-5362]{Hikaru Yabuta}
\affiliation{Hiroshima University \\
Department of Earth and Planetary Systems Science \\
739-8526 Hiroshima, Japan}

\author[0000-0001-6076-3614]{Seiji Sugita}
\affiliation{University of Tokyo \\
Department of Earth and Planetary Science \\
113-0033 Tokyo, Japan}

\begin{abstract}

We analyze images of a rock on Ryugu acquired {\em in situ} by MASCam, camera of the {\sc mascot} lander, with the aim of identifying possible carbonaceous chondrite (CC) analogs. The rock's reflectance ($r_{\rm F} = 0.034 \pm 0.003$ at phase angle $4.5^\circ \pm 0.1^\circ$) is consistent with Ryugu's average reflectance, suggesting that the rock is typical for this asteroid. A spectrophotometric analysis of the rock's inclusions provides clues to CC group membership. Inclusions are generally brighter than the matrix. The dominant variation in their color is a change of the visible spectral slope, with many inclusions being either red or blue. Spectral variation in the red channel hints at the presence of the 0.7~µm absorption band linked to hydrated phyllosilicates. The inclusions are unusually large for a CC; we find that their size distribution may best match that of the Renazzo (CR2) and Leoville (CV3) meteorites. The Ryugu rock does not easily fit into any of the CC groups, consistent with the idea that typical Ryugu-type meteorites are too fragile to survive atmospheric entry.

\end{abstract}



\section{Introduction} \label{sec:intro}

On 3~October 2018, {\sc mascot} was released by the Hayabusa2 spacecraft to land on the small near-Earth asteroid 162173~Ryugu \citep{H17}. Hayabusa2 characterized Ryugu as a dark, moderately hydrated rubble pile, the product of the violent disruption of an undifferentiated, aqueously altered parent body \citep{K19,Su19}. Data acquired by {\sc mascot} during its 17-hour long mission revealed the surface of this C-type asteroid in unprecedented detail. Three scientific instruments collected valuable data. The MASCam camera acquired a total of 120~images, of which 65 show Ryugu's surface \citep{J19}. Images acquired during the descent show rocks and boulders of diverse morphology but no deposits of fine-grained material. MASCam was equipped with LEDs in four colors, covering the visible wavelength range to the near-infrared, to allow imaging at night. Close-up night-time imaging of a small rock in front of the lander revealed abundant multi-colored inclusions set in a dark matrix that resemble those in chondritic meteorites. MASCam images of the landing site allowed the construction of a digital terrain model (DTM) \citep{S19a}. The {\sc mara} radiometer had a good view of the rock imaged by MASCam and determined a low thermal inertia, consistent with a high porosity and low tensile strength \citep{G19}. The {\sc mara} data are incompatible with the presence of an optically thick dust layer on the surface of the rock. The rock's low strength suggests that a Ryugu-type meteorite may not survive atmospheric entry. The Hayabusa2 thermal infrared imager confirmed that the majority of boulders on Ryugu is highly porous, with only a minority being as dense as typical CCs \citep{O20}. The MasMag magnetometer found that Ryugu has no detectable magnetization, which suggests that its parent body did not possess a dynamo \citep{H20}. {\sc mascot}'s fourth instrument, the MicrOmega near-IR spectrometer \citep{B17}, appears not to have acquired useful data. Thus, only MASCam performed an {\em in situ} spectral characterization of Ryugu's surface.

Observations from orbit provided a clear link between the Ryugu surface and carbonaceous chondrite meteorites (CC) \citep{K19,Su19,O20}. The inclusions seen in the MASCam images support the idea that the Ryugu rock is akin to the CCs \citep{J19}. The question is now whether we have meteorite analogs of Ryugu in our collections. \citet{Su19} argued, on the basis of their low reflectance over the visible range, that aqueously altered and thermally metamorphosed meteorites ({\sc atcc}) are good analogs. The Ryugu spectrum has a weak and narrow absorption feature at 2.7~µm that is characteristic for OH-bearing minerals. \citet{K19} found that Ivuna (CI1) heated to $500^\circ$C (i.e.\ an artificial {\sc atcc}) is a good match for Ryugu in terms of overall reflectance and 2.7~µm band strength and shape. \citet{J19} proposed the Tagish Lake meteorite as a possible Ryugu analog. Tagish Lake is a CC of unusually low density that defies straightforward spectral classification; it bears similarities to the CI1 and CM groups, but is distinct from both \citep{Z02}. Its reflectance is unusually low in the visible wavelength range \citep{H01,C12b}, and, unlike typical CI meteorites, it has abundant inclusions \citep{Z02}. All these properties make it an attractive analog candidate.

The present paper extends the preliminary investigation of the Ryugu rock by \citet{J19}, with special focus on the spectrophotometric properties and size distribution of the inclusions. We improved the calibration of the MASCam images, where it proved challenging to properly account for the peculiarities of the LED light source. Our aim is to find the best meteorite analog of Ryugu based on information we can retrieve from the close-up images, i.e.\ inclusion color, abundance, and absolute reflectance. Even if we do not have Ryugu analogs in our collections, we can learn much from a detailed comparison of the Ryugu rock with various carbonaceous chondrite groups. Our analysis also addresses the question of how diagnostic close-up, multispectral images are for the physical characteristics of the asteroid surface. Until we recover the Hayabusa2 samples of Ryugu's surface \citep{S17}, the MASCam images represent the highest resolution observations of C-type asteroid material.

\section{Methods}

\subsection{Data selection}

The analysis in this paper concerns night-time images that show the surface of Ryugu in reflected LED light. {\sc mascot} spent two nights on Ryugu. Images from the first night only show the night sky, but images from the second night show a rock in clear detail. Table~\ref{tab:timeline} lists all image sets acquired by MASCam during these two nights. Five image sets (\#11-15) were acquired in the second night. The first set (11) was taken at dusk and shows rocks in the background still illuminated by the Sun. The last set (15) was taken at early dawn and some images show evidence of daylight in the top right corner, a potential source of stray light. The three remaining sets (12, 13, and 14) are suitable for spectrophotometric analysis. These sets consist of six images: one bias image, one dark image, and one image for each LED color (Table~\ref{tab:set14}). A bias image has the smallest available exposure time of 0.2138~msec. Set~14 is of especially good quality because it was acquired at subzero temperatures, for which detector dark current was negligible \citep{J17}.

\begin{figure}
	\centering
	\includegraphics[width=\textwidth,angle=0]{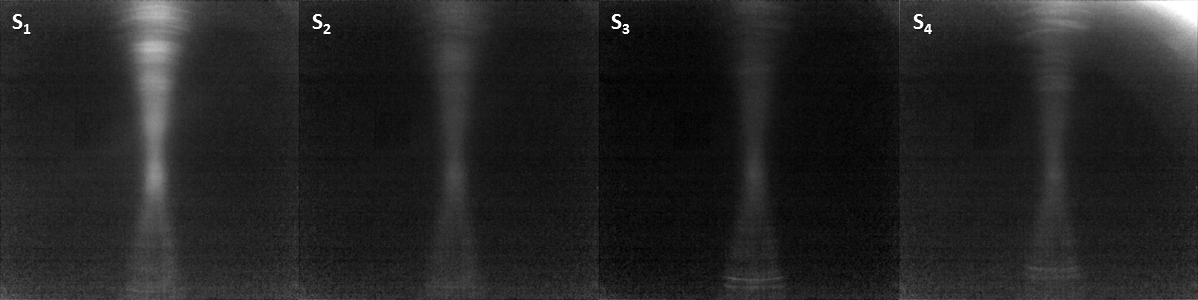}
	\caption{Stray light patterns for all LED colors, shown with identical brightness stretch (black is zero signal).}
	\label{fig:stray_light}
\end{figure}

The images of the empty sky from the first night were expected to be mostly devoid of signal, but instead show stray light that possibly originates internally to the optics from the LEDs. It was not noticed before launch, but was subsequently identified in images acquired on ground. We used the images from the first night to construct stray light patterns for the purpose of calibration. First, we calibrated the images from sets~7 and 8 by subtracting bias and dark current and applying the non-linearity correction \citep{J17}. The bulk dark current was very low, but the presence of many relatively hot pixels made a correction necessary. We then constructed a stray light pattern for each LED color as the average of the two calibrated images (one from either set). The final stray light patterns shown in Fig.~\ref{fig:stray_light} have been convolved by a Gaussian kernel sized $9 \times 9$~pixels with a full-width-at-half-maximum of 3~pixels to reduce noise. The patterns show that stray light in the image center is strongest for the blue LED, and that the infrared LEDs cause strong stray light in the upper right corner.

\subsection{Radiometric calibration}
\label{sec:calibration}

\begin{figure}
	\centering
	\includegraphics[width=\textwidth,angle=0]{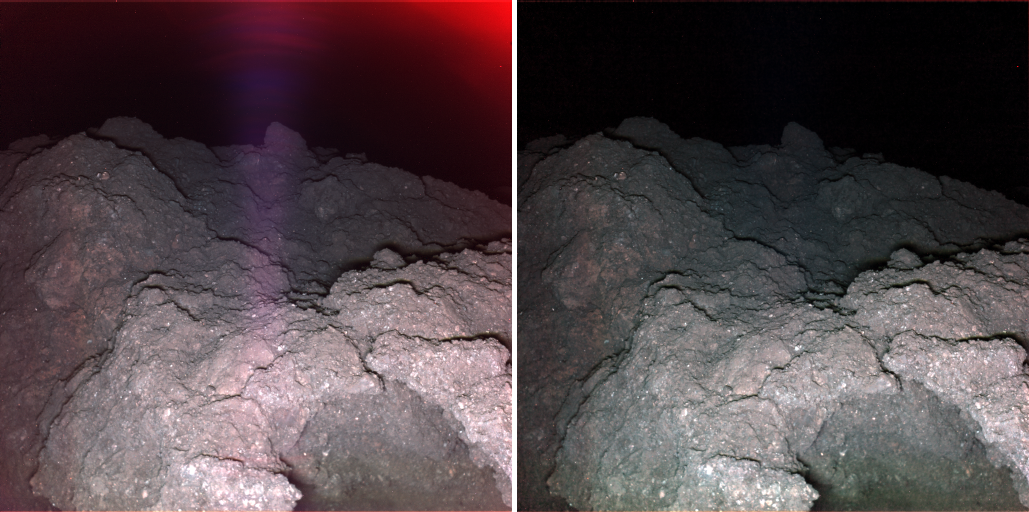}
	\caption{LED stray light correction for the (red, green, blue) = $({\mathbf I}, {\mathbf G}, {\mathbf B})$ composite of set~14. {\bf Left:} Without correction. {\bf Right:} With correction. The brightness of both composites is enhanced identically.}
	\label{fig:correction}
\end{figure}

The LED images of sets~13, 14, and 15 (all numbered \#751-754) were calibrated as described in \citet{J17} with a few modifications. The first step is to produce a clean image by subtracting the bias image (\#755) and applying the non-linearity correction \citep{J17}. While a dark image was acquired (\#750), we omitted the step of correcting for dark current because the low temperature made dark current negligible. We verified that the dark current image contained only noise, and subtracting it in an effort to correct for dark current merely increased the noise in the calibrated image. The resulting clean images ${\mathbf C}_i$ (images are denoted in bold) were calibrated to radiance in [W~m$^{-2}$~sr$^{-1}$] as
\begin{equation}\label{eq:intensity}
{\mathbf L}_i = \frac{{\mathbf C}_i / t_{\rm exp} - {\mathbf S}_i}{R_i {\mathbf V}_i},
\end{equation}
with the exposure time $t_{\rm exp}$ in msec, responsivity factors $R_i$ in [m$^2$~sr~mJ$^{-1}$], and color ratio frames ${\mathbf V}_i$ (see below). Index $i$ indicates the LED color (Table~\ref{tab:responsivities}). In this paper, we refer to $({\mathbf L}_1, {\mathbf L}_2, {\mathbf L}_3, {\mathbf L}_4)$ as $({\mathbf B}, {\mathbf G}, {\mathbf R}, {\mathbf I})$, corresponding to the blue, green, red, and infrared radiance images. Ideally, correcting for differences in irradiance between the LEDs of different colors is done when converting radiance to reflectance. But as the LED irradiance critically depends on topography (which is not accurately known over the entire field-of-view), dividing by the ratio frames in Eq.~\ref{eq:intensity} ensures that color composites constructed from radiance images are correctly balanced. The stray light images ${\mathbf S}_i$, described in the previous section, have the unit DN msec$^{-1}$. Figure~\ref{fig:correction} shows that the stray light correction is substantial for the Ryugu images. Because we also identified stray light in older images that were used to derive the responsivity factors $R_i$ in \citet{J17}, we re-derived the factors from those images with stray light subtracted (Table~\ref{tab:responsivities}). The correction is minor, only 2\% at most. The LEDs were regularly monitored during the cruise phase and were found to be stable. The revised factors are listed in Table~\ref{tab:responsivities}. The exposure time of the LED images in sets~13, 14, and 15 was rather long, at almost 3~seconds, as the auto-exposure algorithm \citep{J17} successfully compensated for the relatively large distance to the dark rock. Only a handful of pixels, inside the brightest inclusions, were overexposed. Overexposure is defined as an image pixels having a signal larger than 12.5~kDN, after bias substraction. Because the non-linearity correction fails for such high values \citep{J17}, overexposed pixels were excluded from the analysis. For set~14, the number of overexposed pixels is: 2 in the blue image, 5 in the green image, and 37 in the red image. No pixels were overexposed in the infrared image because of the comparatively low LED flux at this wavelength.

\begin{figure}
	\centering
	\includegraphics[width=\textwidth,angle=0]{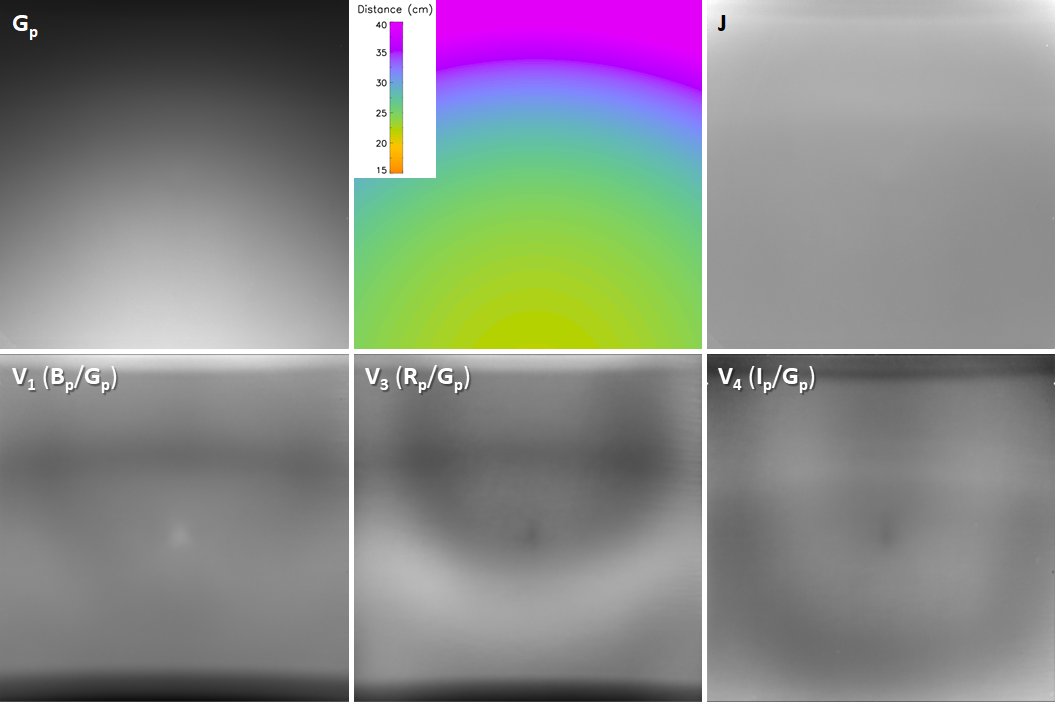}
	\caption{Color ratio frames ${\mathbf V}_i$ were constructed from images of a BaSO$_4$ plate inclined by $30^\circ$ from the horizontal plane. The top row shows the green image  of the illuminated plate (${\mathbf G}_{\rm p}$, black is zero), the distance from the camera aperture to the plate surface, and the irradiance image {\bf J} derived from ${\mathbf G}_{\rm p}$ (black is zero). The bottom row shows the ratio frames, whose brightness is stretched such that $\pm 30$\% of the median is white and black, respectively.}
	\label{fig:ratios}
\end{figure}

The Ryugu rock color composite shown in \citet{J19} features an obvious calibration artifact in the form of a green bar at the bottom. The images in this composite had been calibrated using the color ratio frames ${\mathbf V}_i$ as constructed from images of a barium sulfate plate in the nominal landing configuration, i.e.\ parallel to the bottom side of both {\sc mascot} and the camera. In reality, {\sc mascot} did not come to rest on a flat surface. Fortunately, we had also acquired images of the plate tilted at various angles prior to launch in anticipation of such an occasion. Figure~\ref{fig:ratios} shows the calibrated green image ${\mathbf G}_{\rm p}$ of the plate tilted forward at an angle of $30^\circ$ with respect to the plane parallel to the bottom of the camera (plate images are indicated with the subscript ``p''). This configuration better matches both the distance and orientation of the Ryugu rock with respect to MASCam. Calibrating the plate images included a correction for stray light. The figure also shows a map of the distance to the plate surface, which can be compared to the map of the distance to the Ryugu surface that we will introduce in the following section. We were able to accurately derive the distance to the plate from images of a chessboard pattern that was positioned at the same distance and tilt angle as the plate. From the tilted plate images we constructed the new ratio frames ${\mathbf V}_i$ in Fig.~\ref{fig:ratios}, which are defined with respect to the green image as ${\mathbf V}_1 = {\mathbf B}_{\rm p} / {\mathbf G}_{\rm p}$, ${\mathbf V}_2 = {\mathbf 1}$, ${\mathbf V}_3 = {\mathbf R}_{\rm p} / {\mathbf G}_{\rm p}$, and ${\mathbf V}_4 = {\mathbf I}_{\rm p} / {\mathbf G}_{\rm p}$. The ratio frames are normalized at a location near the image center for which the responsivities in Table~\ref{tab:responsivities} were derived. Division as in Eq.~\ref{eq:intensity} ensures that the barium sulfate plate would appear white in a color composite of radiance images, at least at the location of normalization. In addition, the irradiance {\bf J} in [W~m$^{-2}$] of the green LEDs at unit distance could be determined from ${\mathbf G}_{\rm p}$, where we assumed that the plate has Lambertian reflective properties. For this, we calculated the distance from the center of the LED array to the plate surface and the incidence angle of the light on the plate coming from the same direction, where we used the fact that the LED array is located 1.8~cm below the aperture. The irradiance image {\bf J} in Fig.~\ref{fig:ratios} appears homogeneous over most of the frame, as expected. The irradiance apparently increases towards the top corners, which is likely an artifact resulting from the LED array being extended rather than a point source and/or non-Lambertian reflective properties of the plate. We judge the irradiance image to be reliable in the bottom half of the frame. Knowing the irradiance allows us to estimate Ryugu's absolute reflectance (radiance factor, $I/F$) in image pixel $(x, y)$ as
\begin{equation}\label{eq:reflectance}
(r_{\rm F})^i_{x,y} = \pi \frac{d^2_{x,y} L^i_{x,y}}{d_{\rm ref}^2 J_{x,y}},
\end{equation}
with $d$ the distance from the center of the LED array to the Ryugu surface in cm and reference distance $d_{\rm ref} = 1$~cm.

\begin{figure}
	\centering
	\includegraphics[width=13cm,angle=0]{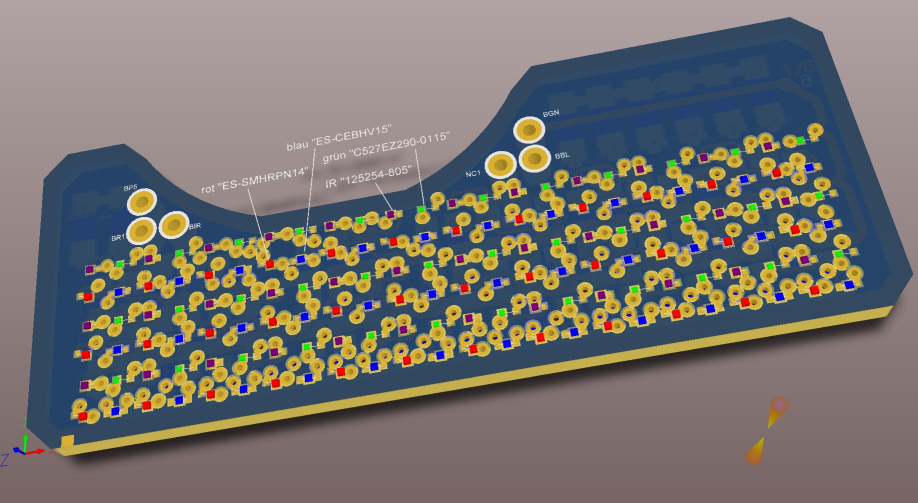}
	\caption{A CAD-drawing of the MASCam LED array showing the distribution of the four colors: blue, green, red, and IR.}
	\label{fig:LED}
\end{figure}

\begin{figure}
	\centering
	\includegraphics[width=\textwidth,angle=0]{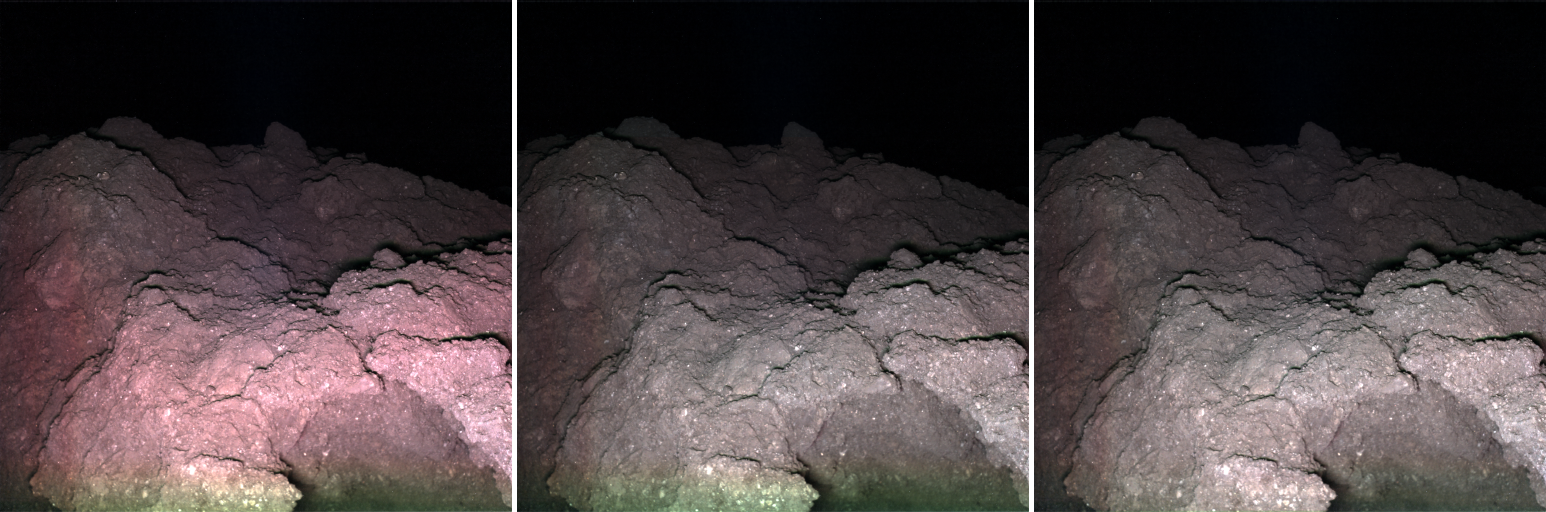}
	\caption{Color correction of the (red, green, blue) = $({\mathbf R}, {\mathbf G}, {\mathbf B})$ composite of set~14 (brightness enhanced). {\bf Left}: No color correction (${\mathbf V}_i = {\mathbf 1}$). {\bf Middle}: Colors corrected using the ${\mathbf V}_i$ in Fig.~19 in \citet{J17}. {\bf Right}: Colors corrected using the ${\mathbf V}_i$ in Fig.~\ref{fig:ratios}.}
	\label{fig:color_cor}
\end{figure}

\begin{figure}
	\centering
	\includegraphics[width=\textwidth,angle=0]{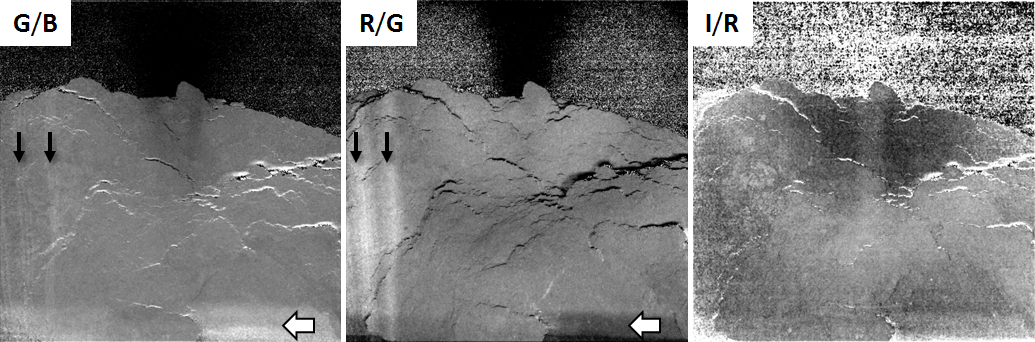}
	\caption{Artifacts in color ratio frames of set~14: Vertical red stripes (black arrows), a horizontal green bar (white arrows), and a dark center for ${\mathbf I} / {\mathbf R}$. The frames are displayed with an similar brightness stretch.}
	\label{fig:artifacts}
\end{figure}

The impact of the color correction (division by the ratio frames ${\mathbf V}_i$ in Eq.~\ref{eq:intensity}) on the calibration to radiance is shown in Fig.~\ref{fig:color_cor}. Here, we compare color composites of uncorrected radiance images, images corrected with the earlier set of ratio frames, and images corrected with the revised frames in Fig.~\ref{fig:ratios}. The revised frames successfully prevent the green bar at the bottom from appearing on the surface close to the camera. The origin of the green bar lies in the obscuration of the bottom row of LEDs on the array by the physical structure surrounding the array. The bottom row has only red and blue LEDs (Fig.~\ref{fig:LED}), and therefore the bar at the bottom of the image for terrain that is close to the camera is green (and IR). Where the terrain is more distant (the apparent hollow at bottom right), the green bar is still faintly visible. This illustrates the difficulty in calibrating images of a surface that is illuminated by this LED device, which has an illumination pattern that is different for each color and, moreover, depends on the distance to the surface. Other artifacts apparently remain in the images after calibration. One way to identify calibration artifacts is to construct ratios of the LED color images. Figure~\ref{fig:artifacts} shows three such ratios: ${\mathbf G} / {\mathbf B}$, ${\mathbf R} / {\mathbf G}$, and ${\mathbf I} / {\mathbf R}$. Both the ${\mathbf G} / {\mathbf B}$ and ${\mathbf R} / {\mathbf G}$ ratios shows traces of the green bar artifact at the bottom of the frame, but only for the more distant terrain. The ${\mathbf G} / {\mathbf B}$ ratio shows a darkening in the center of the frame that may indicate a slight overcorrection of stray light. The ${\mathbf R} / {\mathbf G}$ ratio shows bright vertical stripes at the left of the frame that represent an excess of red signal. This pattern, which had not been seen before, is probably stray light. The stripes are also faintly visible in the ${\mathbf G} / {\mathbf B}$ ratio. The ${\mathbf I} / {\mathbf R}$ ratio frame is relatively dark in the center. We suspect that this is not related to the reflective properties of the rock, but to the illumination pattern of the infrared LEDs being different from that expected (i.e., an incorrect ${\mathbf V}_4$). Contours of rock topography are clearly visible in all ratio frames, either in black or white. This artifact also results from the arrangement of colors over the LED array (Fig.~\ref{fig:LED}). In this case, the terrain was illuminated by the top row of the array (green and IR), while from the perspective of the second row from the top (red and blue) it was in the shadow. This is consistent with the contours showing as white in the ${\mathbf G} / {\mathbf B}$ and ${\mathbf I} / {\mathbf R}$ ratio frames.

\subsection{Inclusion mapping}
\label{sec:mapping}

\begin{figure}
	\centering
	\includegraphics[width=10cm,angle=0]{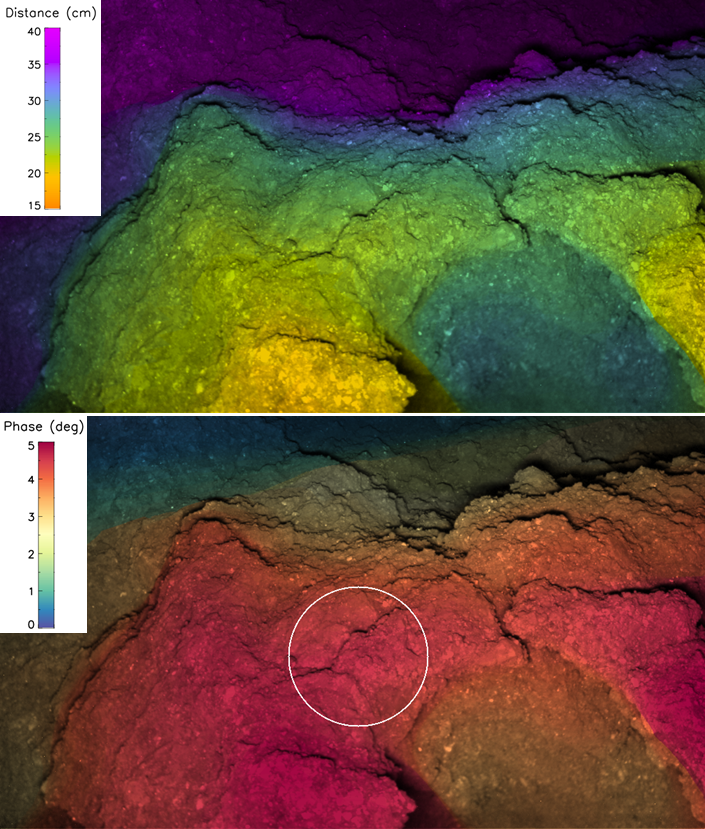}
	\caption{The distance from the camera aperture to the surface ({\bf top}) and illumination phase angle ({\bf bottom}) were calculated from the DTM in \citet{S19a}. The phase angle calculation assumes illumination by a point source at the center of the LED array. The average distance and phase angle inside the circle are $23.6 \pm 1.0$~cm and $4.5^\circ \pm 0.1^\circ$, respectively. The maximum phase angle in the foreground is $5.02^\circ$.}
	\label{fig:shape_model}
\end{figure}

Inclusions were mapped in the radiance $({\mathbf R}, {\mathbf G}, {\mathbf B})$ color composite of set~14 at its full brightness range using the ArcGIS\footnote{\url{https://www.esri.com/en-us/arcgis/about-arcgis/overview}} mapping tools. One of us (HS) did the mapping to achieve consistent results. Inclusions were identified as areas of different brightness or color compared to their immediate surroundings (``matrix''). They were outlined by drawing a polygon between pixels of highest color and brightness contrast at a constant zoom factor. While its vertices were mapped in a continuous coordinate system, the polygon was subsequently converted into a collection of (discrete) image pixels. We did not map inclusions inside inclusions. Identification of inclusions was limited to the well-illuminated terrain in the foreground of the scene, an area we also outlined with a polygon. Inside this area, we defined matrix pixels as all those not associated with inclusions, applying a threshold to the radiance to exclude shadows. The area of each inclusion was calculated from the number of image pixels it covered, the instantaneous field-of-view (IFOV) of those pixels, the distance from the camera aperture to the surface, and the angle between the surface normal and the line connecting aperture and surface (emission angle). As such, the area is corrected to first order for projection effects, as inclusions seen from the side on a sloping surface are larger than they appear. The distance to the rock surface and local emission angle were estimated from the DTM in \citet{S19a}. The color-coded distance in Fig.~\ref{fig:shape_model} shows that most of the rock in the foreground is roughly at 25~cm. The distance to the rock in the background is not known accurately, because of a lack of stereo coverage, but is certainly larger than 40~cm. Mapping was restricted to inclusions on terrain closer than this distance. The surface recedes in an apparent hollow at the bottom right of the scene, and is therefore out-of-focus. The uncertainty of the DTM is about 0.5~cm in the foreground, at a distance of 20~cm, and about 1.5~cm at a distance of 30–40~cm \citep{S19a}. Figure~\ref{fig:shape_model} also shows the phase angle of illumination by the LEDs. The phase angle is generally low, around $5^\circ$ for the well-illuminated surface in the foreground. The low phase angle ensures that we can actually distinguish the inclusions, unobscured by the strong shadows that characterize the day images of this rough surface. The phase angle decreases with distance from the camera, so it is smaller for the poorly-illuminated terrain in the back and the hollow at bottom right. The phase angle was calculated on the assumption of illumination by a point source at the center of the LED array. In reality, any point of the surface received light with a variety of phase angles due to the extended size of the array ($4.2\times 0.9$~cm$^2$). For example, the edges of the array illuminated the terrain in the foreground with a phase angle that was different by about $5^\circ$ from the point source estimate.

Mapping ambiguities arose because of unclear boundaries and spectral variations within apparent inclusions. To illustrate these challenges, we enlarge the area in the foreground at the bottom of the frame in Fig.~\ref{fig:inclusions_closeup}, in which several features are highlighted. Feature~1 appears to be an inclusion with clear boundaries. However, it harbors pixels that appear either red or blue in the color composite, and its average color may therefore appear neutral. Feature~2 is a reddish inclusion whose boundaries are clearly defined in the color composite in (a), but appears to be larger in the stretched composite in (b). Feature~3 was mapped as a collection of three inclusions, one of which is clearly redder than the others. But one could arguably draw an outline around the entire group to count it as a single inclusion. Feature~4 is marked as inclusion on the basis of its slightly brighter appearance, but its boundaries are so indistinct that this feature might as well be an expression of local topography. Finally, the matrix in (b) (i.e.\ pixels that are not marked red) contains abundant small areas that are brighter than their surroundings. In fact, they are so abundant that one suspects that what we have defined as ``matrix'' is not a homogenous substance. These examples make it clear that the ``inclusions'' that we refer to in this report are not uniquely defined.

\section{Results}

\subsection{Rock reflectance}
\label{sec:reflectance}

\begin{figure}
	\centering
	\includegraphics[width=10cm,angle=0]{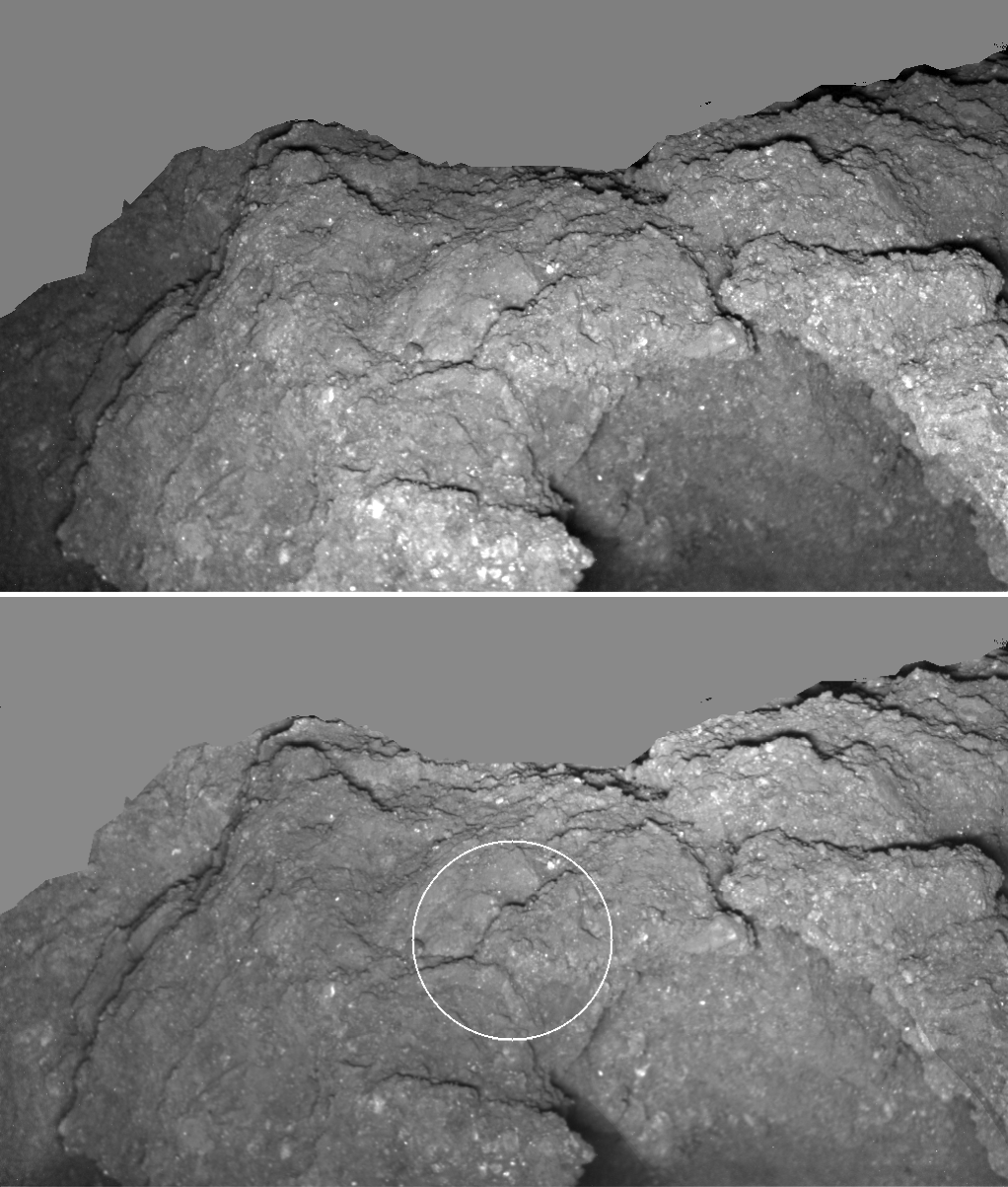}
	\caption{Photometric correction of the green image for terrain closer than 35~cm to the aperture. {\bf Top}: Radiance image {\bf G}. {\bf Bottom}: Reflectance image. The average reflectance inside the circle is $\bar{r}_{\rm F} = 0.034 \pm 0.006$.}
	\label{fig:G_corrected}
\end{figure}

We derived the absolute reflectance ($I/F$) of the Ryugu rock by photometrically correcting the {\bf G} image through Eq.~\ref{eq:reflectance}. Figure~\ref{fig:G_corrected} shows both the uncorrected (radiance) image and the corrected (reflectance) image. Because the DTM is increasingly uncertain beyond 35~cm, we restrict the correction to that distance. The brightness in the reflectance image is much more evenly distributed over the frame than in the radiance image, indicating a successful photometric correction (assuming that the rock reflectance is uniform). For example, the hollow in the foreground is no longer recognizable as such. The reflectance increases towards the terrain in the background, consistent with the lower average phase angle there (see Fig.~\ref{fig:shape_model}). We calculate the average reflectance for the terrain inside the circle in Fig.~\ref{fig:G_corrected} as $\bar{r}_{\rm F} = 0.034 \pm 0.006$ at an average phase angle of $4.5^\circ \pm 0.1^\circ$. The circle was chosen such that it enclosed terrain that is comparatively free of image artifacts (see Fig.~\ref{fig:artifacts}), and is large but small enough such that the phase angle does not vary by too much. While the standard deviation of the reflectance of the pixels inside the circle is 0.006, the true uncertainty of this estimate of the rock reflectance is smaller. It derives from the uncertainty of the green responsivity factor (1\%, Table~\ref{tab:responsivities}), variations in the irradiance image {\bf J} (5\%, Fig.~\ref{fig:ratios}), and uncertainty in the distance to the rock in the DTM. The distance to the area in the circle is about 25~cm (Fig.~\ref{fig:shape_model}), for which the uncertainty is about 1~cm or 4\% \citep{S19a}. Adding up these percentages we arrive at a total uncertainty of 10\%, and estimate the reflectance of the Ryugu rock in the green channel as $r_{\rm F} = 0.034 \pm 0.003$ at $4.5^\circ$ phase angle. We can predict the reflectance of the average Ryugu surface for a phase angle of $4.5^\circ$ using the photometric model parameters from \citet{Su19}, which agree with those of \citet{I14} and are valid for the ONC v-band at $549 \pm 14$~nm \citep{K17}. The predicted $I/F$ values for the $(\iota, \epsilon) = (4.5^\circ, 0^\circ)$, $(2.2^\circ, 2.3^\circ)$, and $(0^\circ, 4.5^\circ)$ geometries are all equal to $0.036$, which agrees very well with our reconstructed $I/F$ of $0.034 \pm 0.003$ at $532^{+12}_{-24}$~nm. Thus, from a photometric point of view the rock in front of {\sc mascot} is typical for Ryugu.

\begin{figure}
	\centering
	\includegraphics[width=10cm,angle=0]{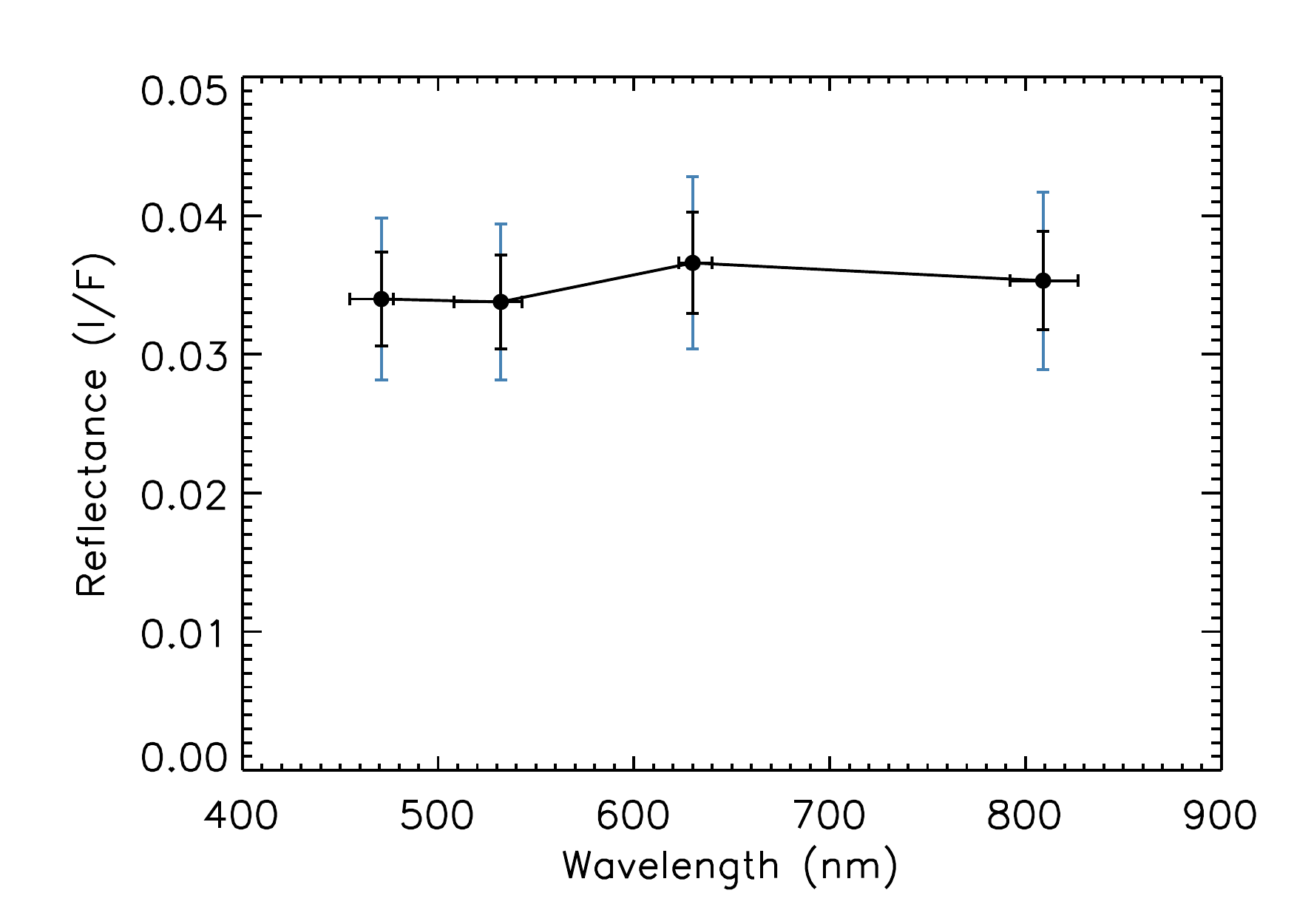}
	\caption{Absolute reflectance of the Ryugu rock, defined as the average of the image pixels inside the circle in Fig.~\ref{fig:G_corrected} and valid for a phase angle of $4.5^\circ \pm 0.1^\circ$. The blue error bars indicate the standard deviation of the pixels, whereas the black error bars provide the uncertainty of the spectral calibration.}
	\label{fig:mean_spectrum}
\end{figure}

The reconstructed reflectance spectrum of the Ryugu rock is shown in Fig.~\ref{fig:mean_spectrum}. The mean reflectance is that of the pixels inside the circle in Fig.~\ref{fig:G_corrected}, with the blue vertical error bars indicating the standard deviation. The sizes of the error bars reflect the brightness variations over the terrain, which includes shadows, but not the uncertainty of the spectral shape. The latter is the uncertainty of the spectral calibration, which derives from the uncertainty of the responsivity factors (only 1\%) and that of the LED irradiance in the different color channels. The (spectral) uncertainty in the LED irradiance dominates the calibration uncertainty, and concerns the color correction, i.e.\ division by the color ratio frames in Fig.~\ref{fig:ratios}. As these frames are scene-dependent, their uncertainty is difficult to quantify. For example, frame ${\mathbf V}_3$ was constructed from images of a flat plate and has large-scale variations of around 15\%. The uncertainty of the color correction is probably smaller than that. We adopted an uncertainty of 10\% for the black vertical error bars on the rock reflectance spectrum in Fig.~\ref{fig:mean_spectrum}. Within this uncertainty, our spectrum is consistent with average reflectance spectra of Ryugu from ONC images \citep{Su19}. The slight excess of the MASCam reflectance in the red channel is not necessarily real, as it is within the calibration uncertainty.

\subsection{Inclusions}

\subsubsection{Spatial distribution}
\label{sec:spatial_dist}

We mapped a total of 1443 inclusions inside an area that was well illuminated due to its proximity to the camera. The totality of this area minus inclusions is defined as matrix, where we applied a lower threshold of 0.008~W~m$^{-2}$~sr$^{-1}$ to the radiance (for all LED colors) to exclude areas in shadow. A map of inclusion and matrix pixels is shown in Fig.~\ref{fig:inclusions_map}. The inclusions appear to be more or less uniformly distributed over the mapping area. We tried to estimate the total area covered by inclusions and matrix to estimate their respective volume abundances, which are thought to be diagnostic quantities. However, we found the total matrix area to be dominated by pixels with an emission angle close to $90^\circ$. The area derived for such pixels is probably very inaccurate. The resolution of the \citet{S19a} DTM appears to be much lower than the actual scale of the surface roughness, so the emission angle calculated for image pixels is generally only a first order approximation. We therefore selected only matrix pixels with an emission angle smaller than $80^\circ$ (excludes 1.4\%). The total area of the matrix pixels is then 172~cm$^2$. For the same reason we selected only inclusions with an average emission angle smaller than $80^\circ$ (excludes 6). The total area of the mapped inclusions is then 13.6~cm$^2$. The areal abundance of inclusions is $100\% \times 13.6 / (13.6 + 172) = 7.3\%$, which corresponds to a matrix areal abundance of 92.4\%. As a test we also estimated the pixel areas without the correction for emission angle and found inclusion and matrix abundances of 7.2\% and 92.8\%, respectively. These values are virtually identical to the earlier ones, implying that the retrieved abundances are robust. However, the terrain we mapped as ``matrix'' may include unresolved inclusions, so our inclusion and matrix abundances are lower and upper limits, respectively. Finally, we performed a simple numerical simulation of small, randomly distributed spheres throughout a volume and verified that the volume abundance (vol.\%) can be estimated as the observed areal abundance on a planar surface. Thus, to first order, the volume abundance is equal to the retrieved areal abundance.

\subsubsection{Spectral properties}

\begin{figure}
	\centering
	\includegraphics[width=10cm,angle=0]{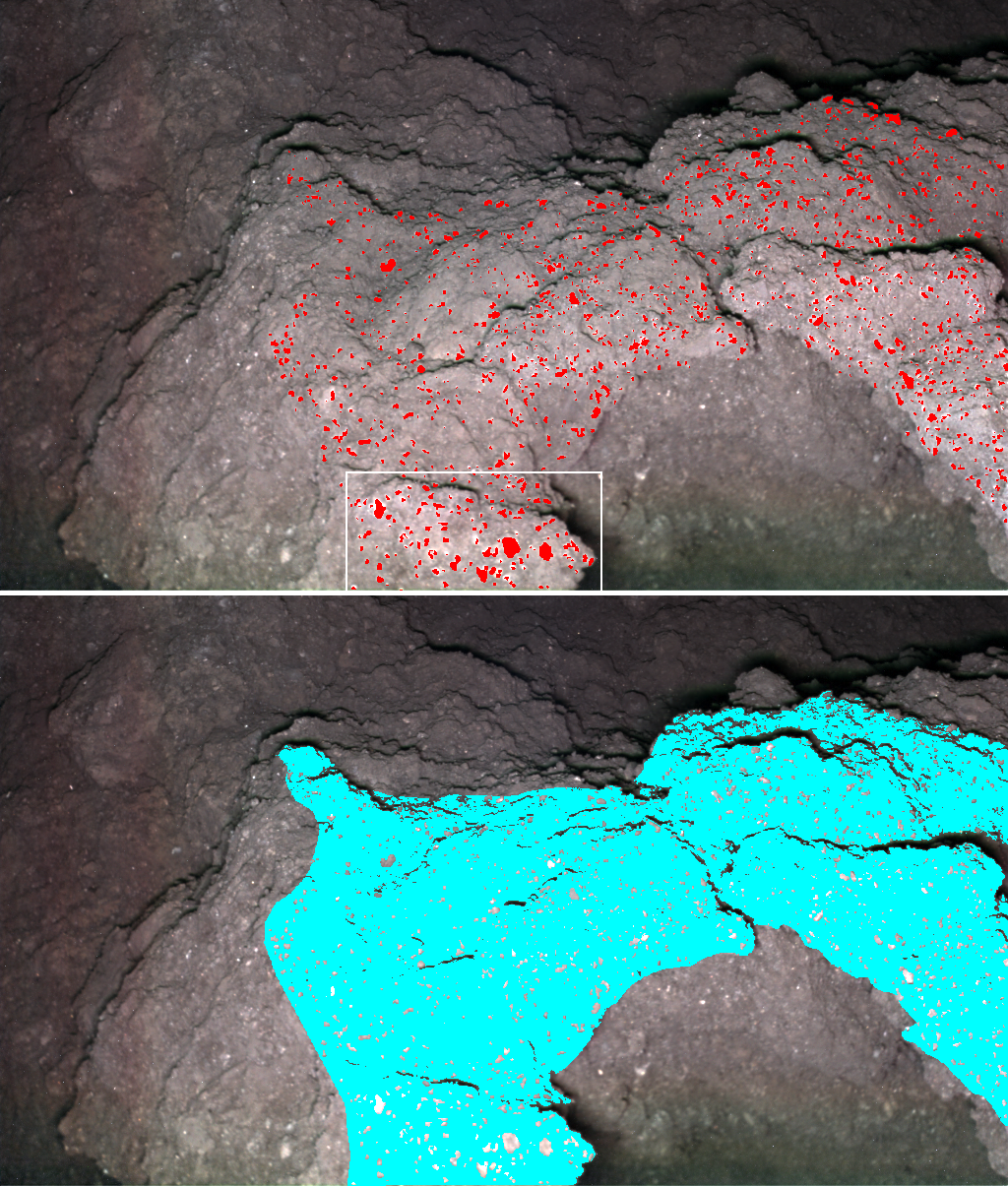}
	\caption{Definition of inclusions and matrix. {\bf Top}: Inclusion pixels (red). The area in the rectangle is enlarged in Fig.~\ref{fig:inclusions_closeup}. {\bf Bottom}: Matrix pixels (cyan).}
	\label{fig:inclusions_map}
\end{figure}

\begin{figure}
	\centering
	\includegraphics[width=8cm,angle=0]{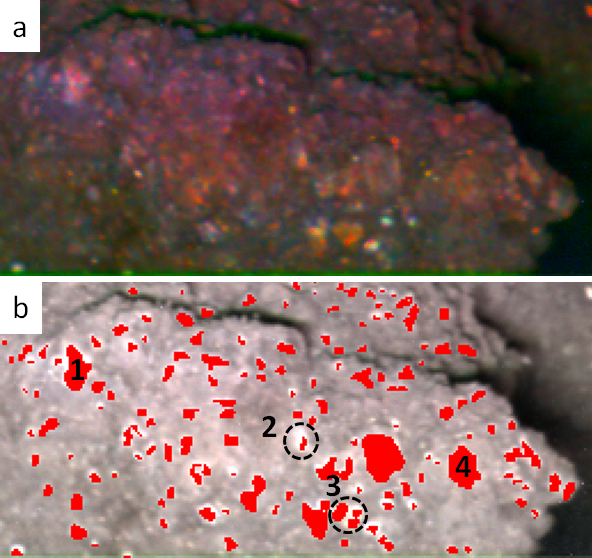}
	\caption{Challenges associated with mapping of inclusions. The area shown is that indicated in Fig.~\ref{fig:inclusions_map}. ({\bf a})~Saturated $({\mathbf R}, {\mathbf G}, {\mathbf B})$ composite at full brightness range with red intensity reduced. ({\bf b})~Corresponding map with inclusions marked in red on a stretched color composite background. Features labeled 1-4 are discussed in the text.}
	\label{fig:inclusions_closeup}
\end{figure}

\begin{figure}
	\centering
	\includegraphics[width=11cm,angle=0]{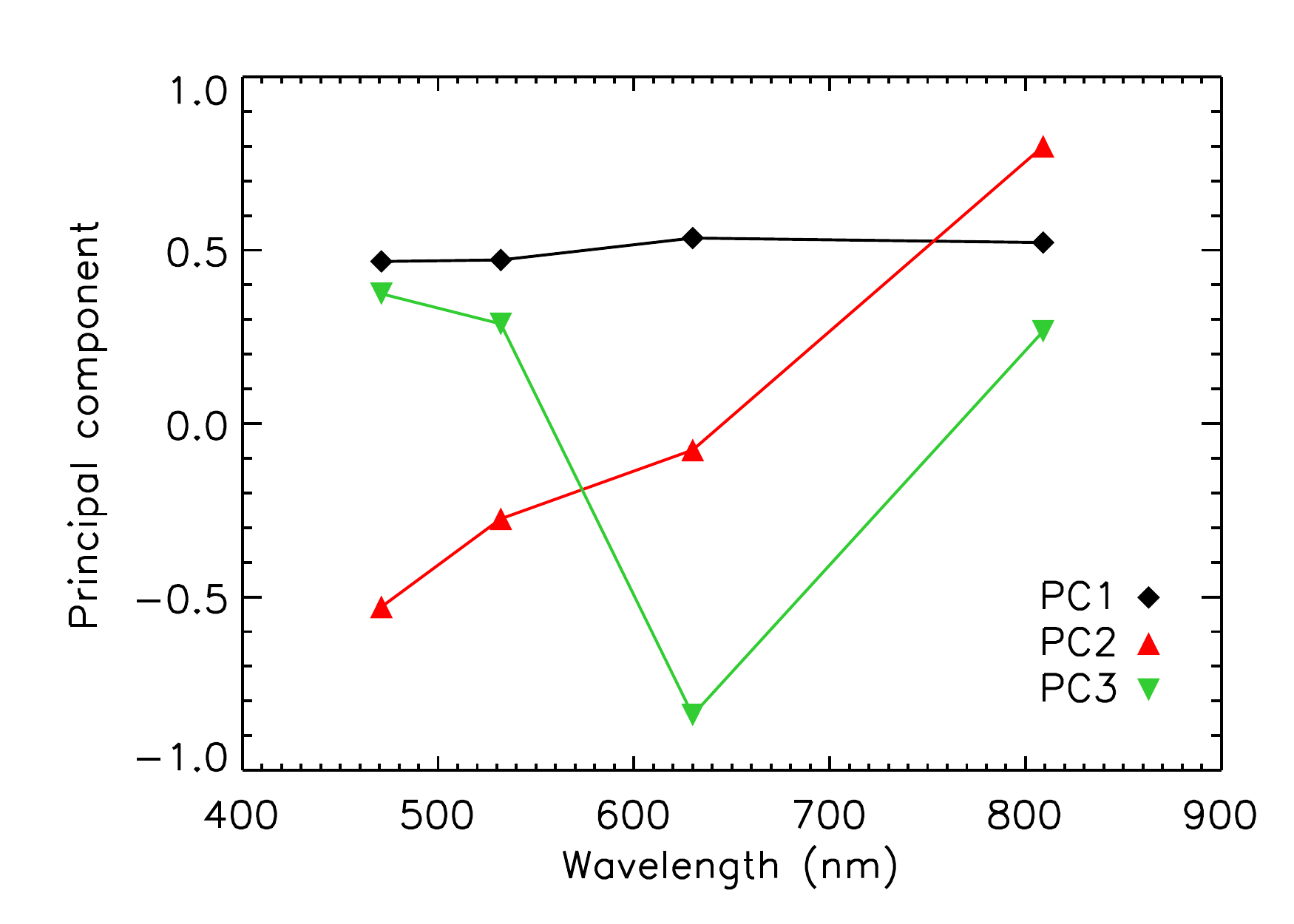}
	\caption{Principal components of the inclusion spectra. The contribution of the three components to the variance is PC1: 97.9\%, PC2: 1.4\%, and PC3: 0.5\%.}
	\label{fig:pca}
\end{figure}

\begin{figure}
	\centering
	\includegraphics[width=11cm,angle=0]{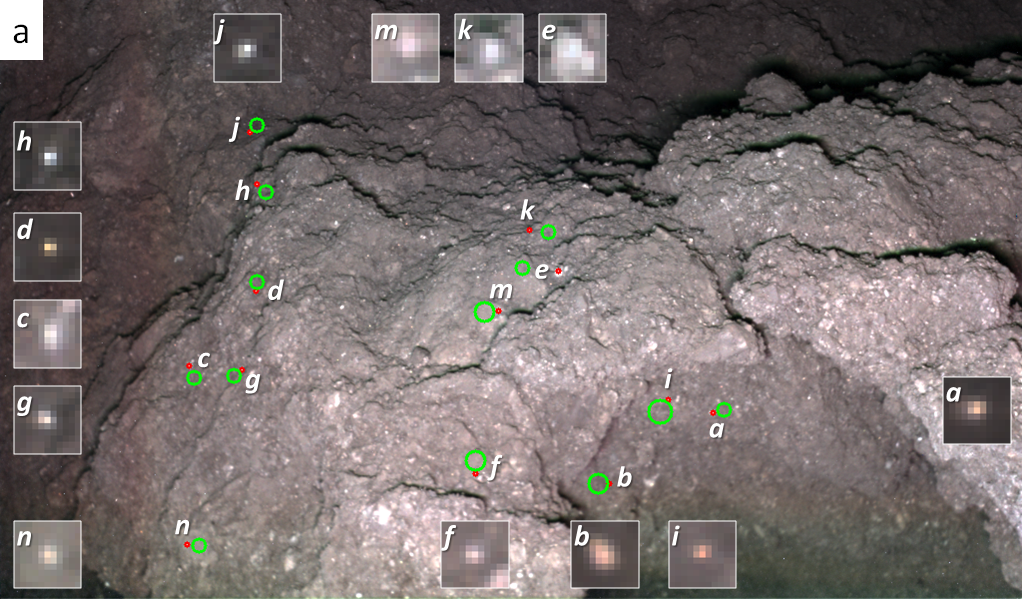}
	\includegraphics[width=12cm,angle=0]{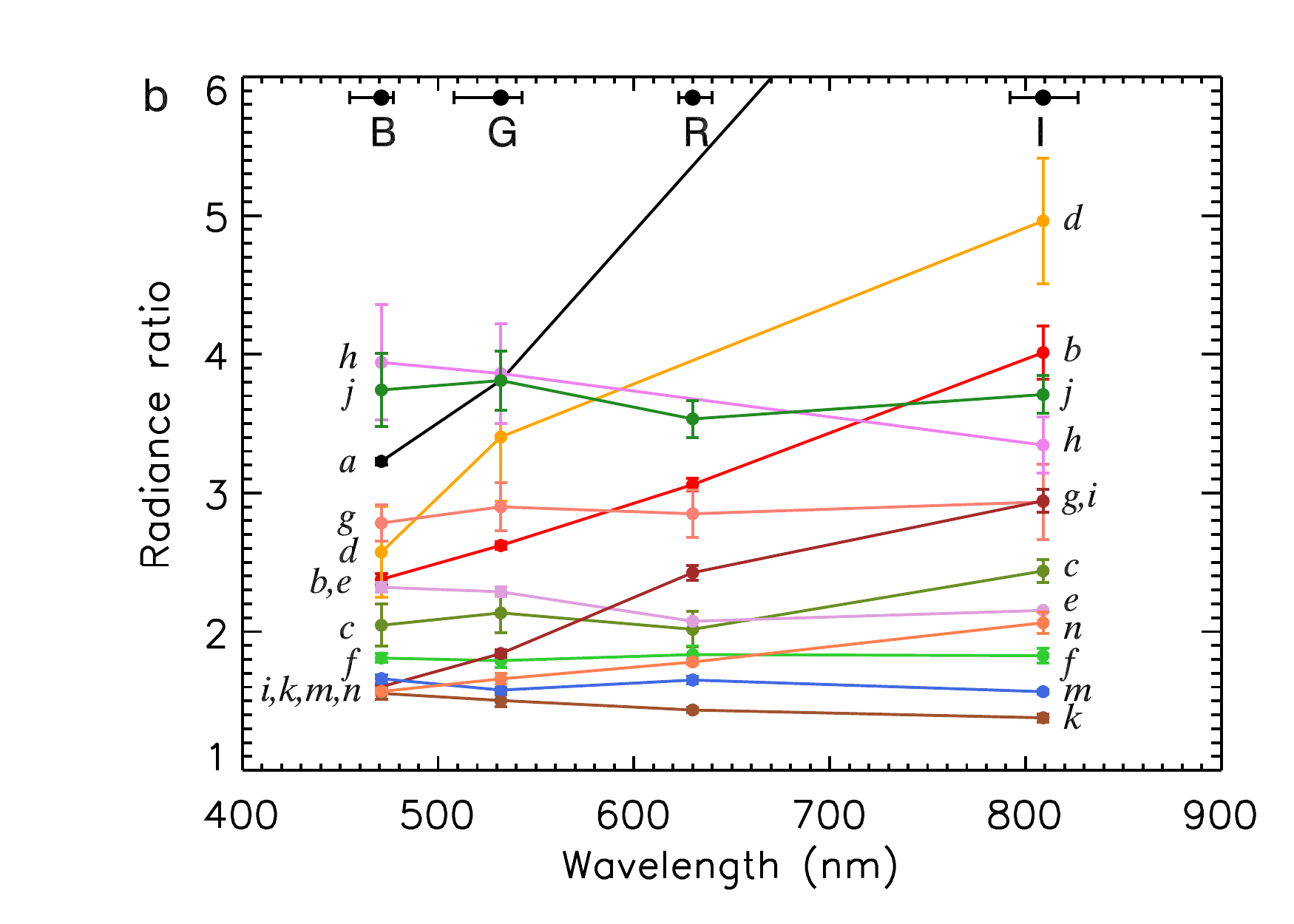}
	\caption{Inclusion spectral diversity. ({\bf a})~Map of selected inclusions, labeled $a$-$n$. The red circles indicate the inclusion locations whereas the green circles indicate nearby featureless matrix areas. The insets are $({\mathbf R}, {\mathbf G}, {\mathbf B})$ composites at full brightness range (zero is black). The inclusion radiance was determined for the pixel at the center of the inset. ({\bf b})~Ratio spectra of all labeled inclusions, calculated as the radiance of the central pixel divided by the average radiance of the associated matrix. Each data point is the radiance ratio for that pixel averaged over the image sets~13, 14, and 15 ($n = 3$), with the error bars indicating the standard deviation. Name and wavelength range of the color channels are indicated at the top. The red (R) data point is missing for some inclusions because of overexposure. The infrared (I) ratio for inclusion~$a$ plots outside the range as $8.2 \pm 0.3$.}
	\label{fig:single_spectra}
\end{figure}

The color of the inclusions can be determined as the ratio of color-corrected radiance images and thus does not depend on an accurate determination of the LED irradiance. We investigate the color of the inclusions as defined in Fig.~\ref{fig:inclusions_map}. We determined the average spectrum of each inclusion and subjected the body of spectra ($n = 1441$) to a principal component analysis (PCA) using the prcomp package in R\footnote{\url{https://www.r-project.org}}. The first three principal components (or eigenvectors) are shown in Fig.~\ref{fig:pca}, where we omitted the last component (PC4) as the highest components generally contain only noise. PC1 represents the average shape of the inclusion spectra, which is similar to the average spectrum of the rock itself (Fig.~\ref{fig:mean_spectrum}). PC2 represents the dominant color variation, which in this case is a change of the spectral slope from blue to infrared, while PC3 expresses more subtle color variations that may exist on top of the slope variation. The contribution of the three components to the variance is PC1: 97.9\%, PC2: 1.4\%, and PC3: 0.5\%. The large contribution of PC1 is the consequence of the variable degree of illumination over the scene, where inclusions in the background are perceived as darker than those in the foreground. The dominant spectral variation (PC2) is therefore a change in spectral slope over the entire wavelength range of the LEDs. PC3 uncovers a possible variation in the red channel with respect to the other three, which may be related with red being slightly depressed in PC2. To better understand this variation, we evaluate the spectra of several individual inclusions. Many inclusions have a single pixel at their center that is much brighter than the others, indicating that they are unresolved, in which case the image merely represents the point-spread-function of the imaging system. We therefore focus only on that brightest pixel. To circumvent the uncertainties associated with unequal illumination patterns between the LEDs, we calculate the ratio of the radiance in the brightest pixel and the average radiance of the matrix, chosen as a (circular) area of uniform appearance in close proximity. Figure~\ref{fig:single_spectra} shows the ratio spectra of 14~prominent inclusions, labeled $a$-$n$. Some of these are outside the terrain mapped earlier, as, in this case, the selection criterion is radiometric accuracy (lack of image artifacts) rather than DTM accuracy. We calculated ratio spectra for each of the image sets~13, 14, and 15 (Table~\ref{tab:timeline}), and the ratio spectra in the figure are averages over these three sets. A few inclusions ($a$, $b$, $d$) are very bright and very red. Inclusion~$a$ is the most extreme case, being more than 8~times brighter than the matrix in the infrared. Another inclusion ($h$) is also much brighter (by about a factor~4) but blue instead. Other inclusions are typically twice as bright as the matrix, and either red, blue, or spectrally neutral. Variation in the red channel indeed exists. The radiance in the red channel appears to be reduced for inclusions $c$, $e$, and $j$, although the error bars are relatively large, and may be slightly enhanced for inclusions $i$ and $m$.

\begin{figure}
	\centering
	\includegraphics[width=11cm,angle=0]{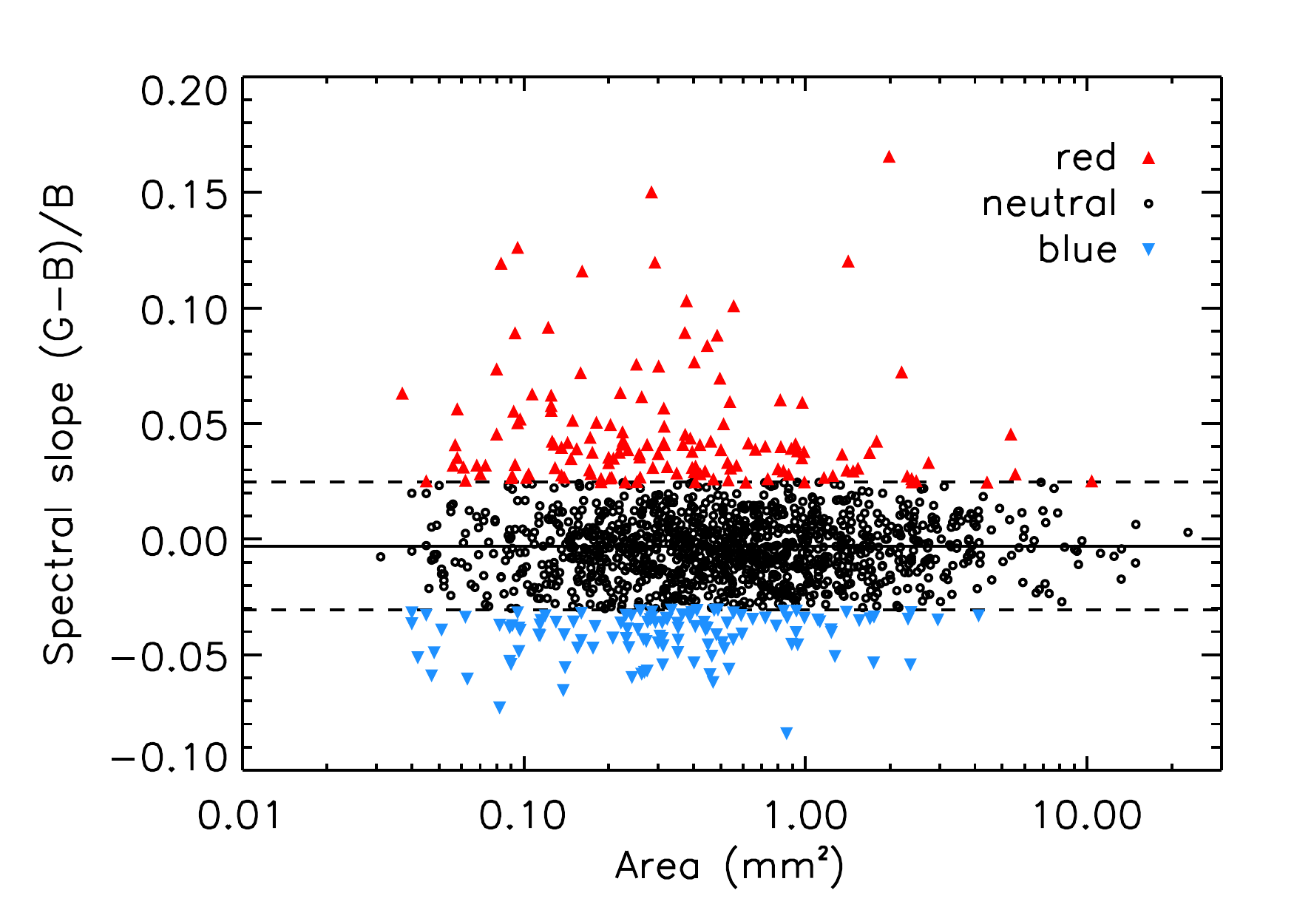}
	\caption{The relative spectral slope, defined as $({\mathbf G} - {\mathbf B}) / {\mathbf B}$, of all mapped inclusions as a function of inclusion area. The drawn line is the average spectral slope of the matrix pixels, with the dashed lines indicating the standard deviation. We define ``red'' inclusions as those with a spectral slope larger than the upper dashed line, and ``blue'' inclusions as those with a slope smaller than the lower dashed line.}
	\label{fig:spectral_slope}
\end{figure}

We return to the inclusions as they are mapped in Fig.~\ref{fig:inclusions_map}. We express the dominant color variation with a single quantity: the relative spectral slope. In light of the lack of apparent artifacts in the mapping area in the ${\mathbf G} / {\mathbf B}$ ratio frame in Fig.~\ref{fig:artifacts} and the ``flat'' appearance of the rock, we define the relative slope as $({\mathbf G} - {\mathbf B}) / {\mathbf B}$. Figure~\ref{fig:spectral_slope} shows the relative slope as a function of inclusion area. The matrix pixels are neutral in color, with a relative spectral slope of $-0.003 \pm 0.028$. We now define ``red'' inclusions as those with a spectral slope larger the average spectral slope of the matrix pixels plus one standard deviation, and ``blue'' inclusions as those with a slope smaller than the average matrix slope minus one standard deviation. The figure shows that the vast majority of inclusions have spectral slopes that are in the range of those of the matrix pixels. This suggests that the matrix harbors unresolved or otherwise unrecognized inclusions. Also, some inclusions harbor both red and blue pixels (examples in Fig.~\ref{fig:inclusions_closeup}), which tends to reduce their average spectral slope to that of the matrix. The number of red inclusions is 1.2~times larger than the number of blue inclusions. This ratio is within the expected uncertainty assuming Poisson statistics, so the number of red and blue inclusions is not significantly different. However, the ratio of red over blue inclusions critically depends on the definition of spectral slope: Had we defined the slope as $({\mathbf R} - {\mathbf B}) / {\mathbf B}$ or $({\mathbf I} - {\mathbf B}) / {\mathbf B}$, the ratio would be 1.4 and 4.5, respectively. Figure~\ref{fig:spectral_slope} also reveals that inclusion color is not a matter of size, although the largest inclusions in our sample ($> 3$~mm$^2$) are strictly neutral in color, perhaps as a result of averaging any spectral diversity inside. Finally, we counted 9~very red inclusions (slope $>0.1$), but none that are similarly blue (slope $<-0.1$).

\subsubsection{Size distribution}
\label{sec:size}

\begin{figure}
	\centering
	\includegraphics[width=11cm,angle=0]{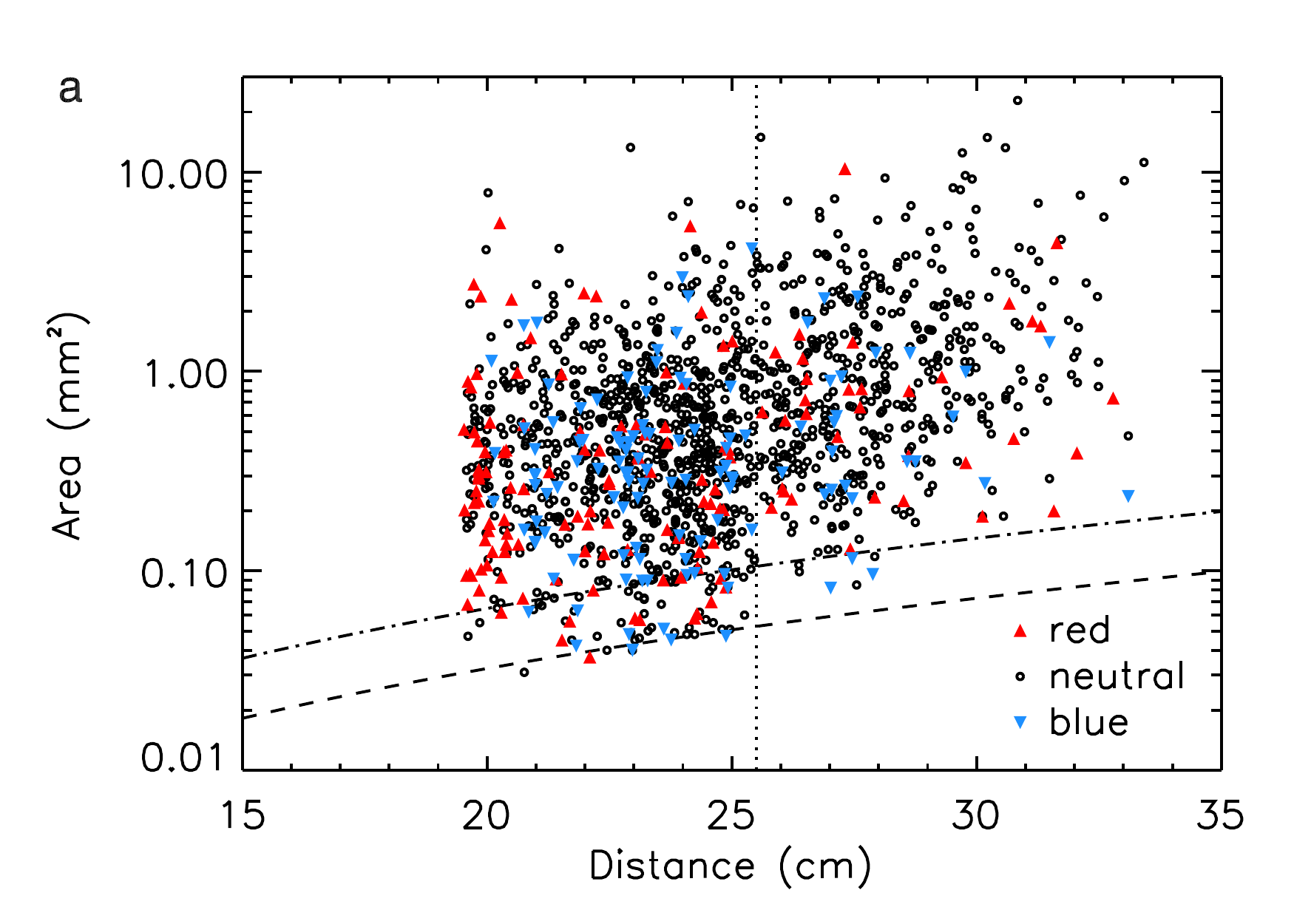}
	\includegraphics[width=11cm,angle=0]{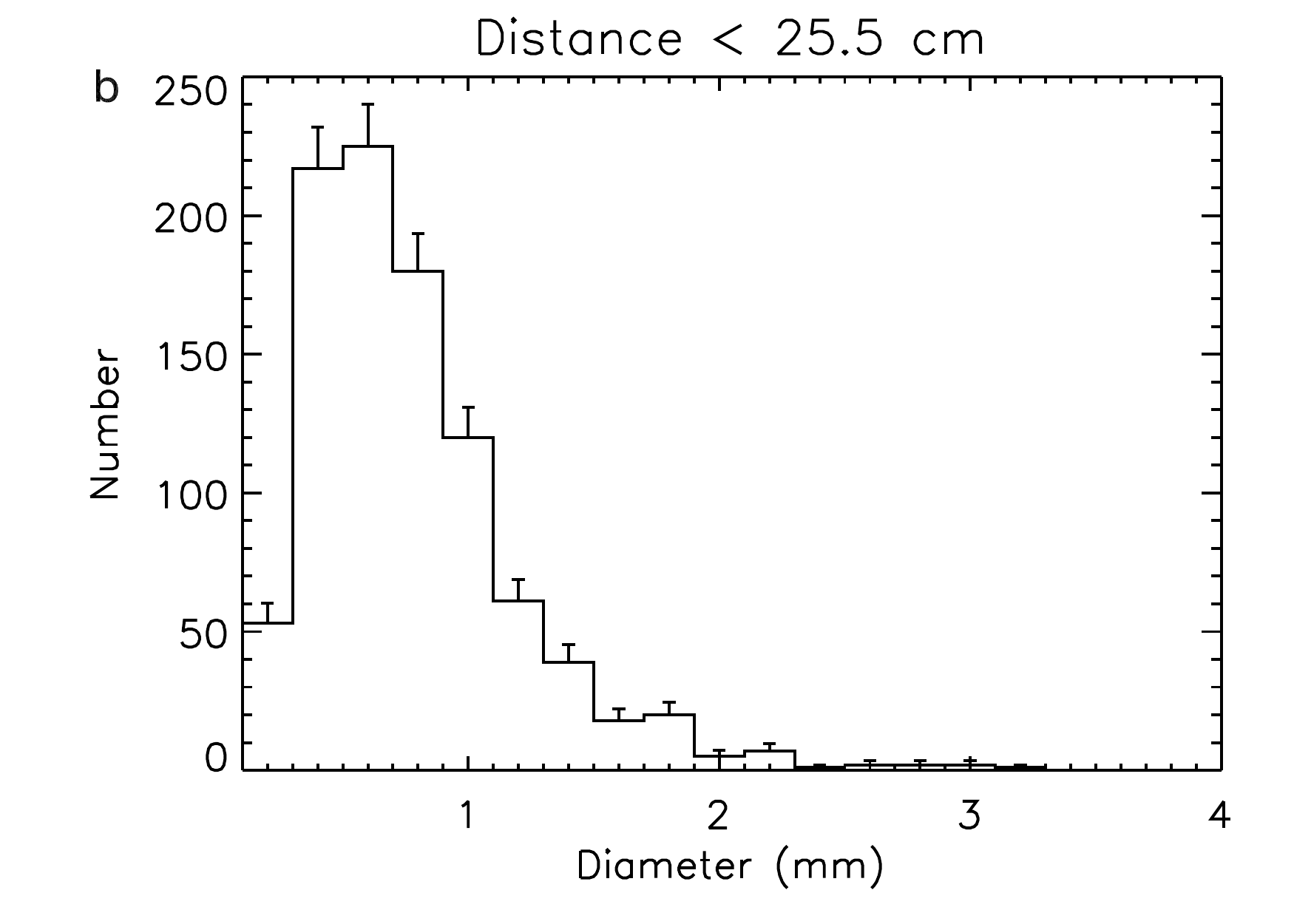}
	\caption{Inclusion size distribution, with inclusion color defined in Fig.~\ref{fig:spectral_slope}. ({\bf a})~Projected inclusion area as a function of distance from the aperture. The dashed line corresponds to the area covered by a single pixel with a IFOV of 0.9~mrad, whereas the dash-dotted line corresponds to the area covered by two such pixels. The space between the two lines represents a sampling gap. Inclusions left of the dotted line were selected for the histogram. ({\bf b})~Histogram of the inclusion diameter, defined for an disk of identical area, with Poisson error bars. The counts only include inclusions on terrain closer than 25.5~cm ($n = 954$). There is 1~inclusion with a diameter larger than 4~mm.}
	\label{fig:inclusion_area}
\end{figure}

The smallest and largest inclusions in our sample have an area of 0.031 and 23~mm$^2$, respectively. The size distribution of the mapped inclusions is affected by three biases. Bias~(1) is that the smallest inclusions can only be distinguished on terrain closest to the camera. Figure~\ref{fig:inclusion_area}a shows the inclusion area as a function of distance to the camera aperture. It shows that inclusions were mapped on terrain at a distance between 19 and 34~cm. The dashed curve represents the area of a single pixel with an IFOV of 0.9~mrad seen face-on (zero emission angle) as a function of distance. The smallest inclusions cluster around this curve, indicating that their size is a single pixel. We cannot be sure that such inclusions are fully resolved, so their area is an upper limit. Another curve (dash-dotted) represents inclusions that cover an area of two image pixels. The mostly empty space between the curves is a sampling gap. Similar gaps exist for higher pixel numbers. Bias~(2) is introduced by the sloping terrain at larger distance; the higher emission angles there make it impossible to see many small inclusions. Bias~(3) follows from the fact that most of the area we mapped is at larger distance, so there we find a relatively large number of large inclusions. The three kinds of bias make it difficult to evaluate whether inclusions are uniformly distributed over the rock, i.e.\ independent of their size. At least when we distinguish red and blue inclusions, again defining the relative slope as $({\mathbf G} - {\mathbf B}) / {\mathbf B}$, we find little variation; both red and blue inclusions are found at any distance. However, if we define the relative spectral slope as $({\mathbf R} - {\mathbf B}) / {\mathbf B}$, there appear to be more blue than red inclusions at larger distance. The distribution is uniform again for $({\mathbf I} - {\mathbf B}) / {\mathbf B}$, so the odd result for the red channel may reflect heterogeneity in the rock but could also be an artifact.

The three biases make it necessary to restrict the size analysis to inclusions mapped on terrain relatively close to the camera. It appears that single-pixel-sized inclusions were only recognized up to a distance of 25.5~cm (dotted vertical line in Fig.~\ref{fig:inclusion_area}a), so we adopt this as a distance limit. Inclusion size is often reported in the literature as the diameter of a sphere. We therefore convert our inclusion area $a$ into diameter $d = 2 \sqrt{a / \pi}$ of an equivalent disk. Figure~\ref{fig:inclusion_area}b shows a histogram of the diameter of inclusions closer than 25.5~cm. The minimum and maximum diameter in the sample are 0.20 and 5.4~mm, respectively. The vast majority of inclusions is smaller than 1~mm in diameter, and very few are larger than 2~mm. The histogram peaks around 0.5~mm diameter. However, the number in the smallest size bin (0.1-0.3~mm) is uncertain for two reasons. First, the area of a single pixel with an IFOV of 0.9~mrad seen face-on at 20~cm distance corresponds to $d = 0.20$~mm. So some inclusions assigned to this bin could actually be smaller than 0.1~mm, but so bright that they appeared bigger due to the detector's point spread function (PSF). On the other hand, the rock displays many small, faint brightness features (see examples in Fig.~\ref{fig:inclusions_closeup}), which we assumed to be photometric variations resulting from topography, but might also be inclusions of the correct size that were too faint to be recognized. As there are competing biases for the number in the smallest size bin, it is probably not too far off, and the peak in the size distribution around 0.5~mm is likely real.

\subsubsection{Brightness distribution}
\label{sec:brightness}

\begin{figure}
	\centering
	\includegraphics[width=11cm,angle=0]{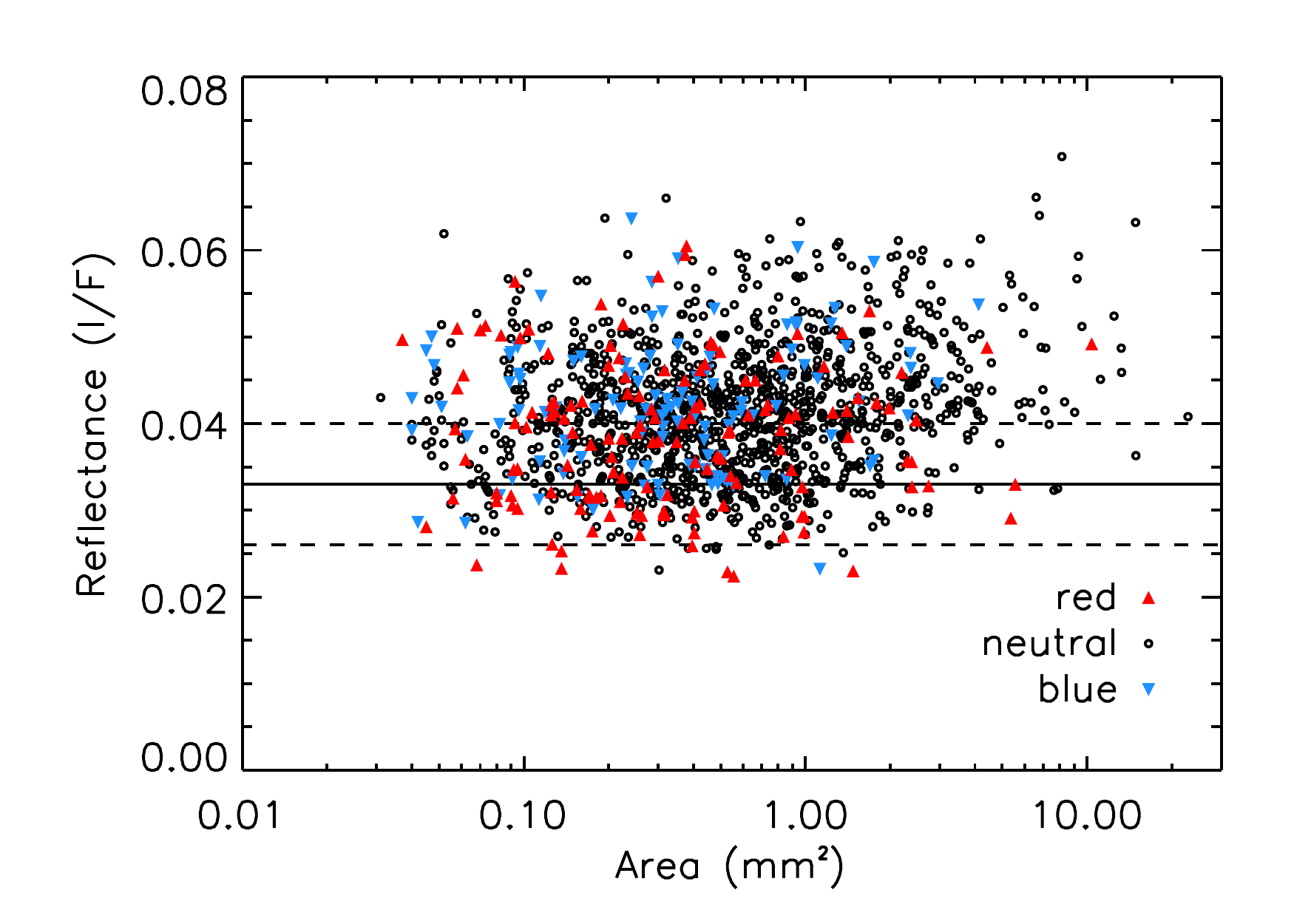}
	\caption{Absolute reflectance of the inclusions derived from the photometrically corrected ${\mathbf G}$ image, with inclusion color as defined in Fig.~\ref{fig:spectral_slope}. The drawn line is the average reflectance of the matrix pixels, with the dashed lines indicating the standard deviation. The average phase angle of all inclusions is $4.6^\circ \pm 0.5^\circ$.}
	\label{fig:inclusion_reflectance}
\end{figure}

The successful photometric correction described in Sec.~\ref{sec:reflectance} allows us to reconstruct the absolute reflectance of the inclusions. Inclusions generally appear bright in the images, and we did not unequivocally identify any that are darker. Figure~\ref{fig:inclusion_reflectance} compares the reflectance of the inclusions with the average matrix reflectance, again distinguishing red and blue inclusions as defined in Fig.~\ref{fig:spectral_slope}. The figure confirms our impression that inclusions are generally brighter, by up to a factor of two, than the matrix at phase angles of around $5^\circ$. The brightness of inclusions seems to depend on neither color nor size.

\section{Discussion}

In the previous sections we determined the reflective properties of the Ryugu rock and its inclusions and derived the size distribution for the latter. Our results suggest that the rock is typical for Ryugu, although this should be confirmed by an analysis of Hayabusa2 observations of the landing site\footnote{A paper on this topic is in preparation.}. We should now be in a position to choose the most appropriate carbonaceous chondrite (CC) analog group for Ryugu rocks from the perspective of MASCam. The low albedo over the entire visible wavelength range poses a challenge to match Ryugu with any CC group \citep{Su19}, so we will search for clues in the inclusion size distribution and their spectrophotometric properties.

Inclusions in the Ryugu rock appear to be exclusively brighter than the surrounding matrix. A principal component analysis reveals that their dominant spectral variability is a variation in visible spectral slope, with an additional, more subtle, spectral variation existing around 0.63~µm, as expressed by PC3 (Fig.~\ref{fig:pca}). The shape of our PC3 is reminiscent of that of PC3 in a PCA of visible spectra of CCs in \citet{Hi17}. The authors linked a depression in their PC3 at 0.7~µm to variability in the 0.7~µm absorption band, which is attributed to Fe$^{2+}$-Fe$^{3+}$ charge transfer in hydrated phyllosilicates like serpentines and saponites \citep{VG89,C11a}. It is tempting to attribute the shape of our PC3 to the presence of this absorption band in the spectrum of some inclusions, but there are several complicating factors. First, the higher PCs are usually affected, or even dominated, by noise. Second, the \citet{Hi17} PCA was for a variety of carbonaceous chondrites, whereas ours is for a variety of inclusions inside a single, putative carbonaceous chondrite. Third, the MASCam red channel is centered at 0.63~µm, which may not be deep enough into the (broad) 0.7~µm band for a positive identification. Confirming the presence of the 0.7~µm band in inclusion spectra would have important implications. The spectrum of the rock as a whole is consistent with that of average Ryugu (Fig.~\ref{fig:mean_spectrum}). It does not show this band, meaning that the matrix does neither. \citet{Su19} and \citet{K19} argued that the closest meteorite analog for Ryugu are heated CCs. If some inclusions display the 0.7~µm band, the implication of their argument is that the rock matrix was heated, but the inclusions were not. Such a sequence of events is difficult to envision.

Ideally, we put these results in context of other studies of CC inclusion color variability in the visible wavelength range. Unfortunately, such studies appear to be scarce, probably because spectral features diagnostic for the composition are generally located in the near-IR. Also, spectral studies commonly deal with CC powders, but the images of the Ryugu rock are more akin to those of meteorite fragments or slabs. Reflectance spectra of CC slabs may be more blue-sloped and generally darker than spectra of powders \citep{C18}. MASCam found Ryugu inclusions to be either blue (negative spectral slope) or red (positive slope). Enrichment in olivine appears to give rise to a blue spectral slope in the visible for bright inclusions in the Murchison meteorite \citep{G15}, which may also explain the color of blue Ryugu inclusions. Many Ryugu inclusions are red, some of them very red, which may hint at the presence of phyllosilicates, Fe-rich oxides, or spinel \citep{C11b,G15}.

\begin{figure}
	\centering
	\includegraphics[width=11cm,angle=0]{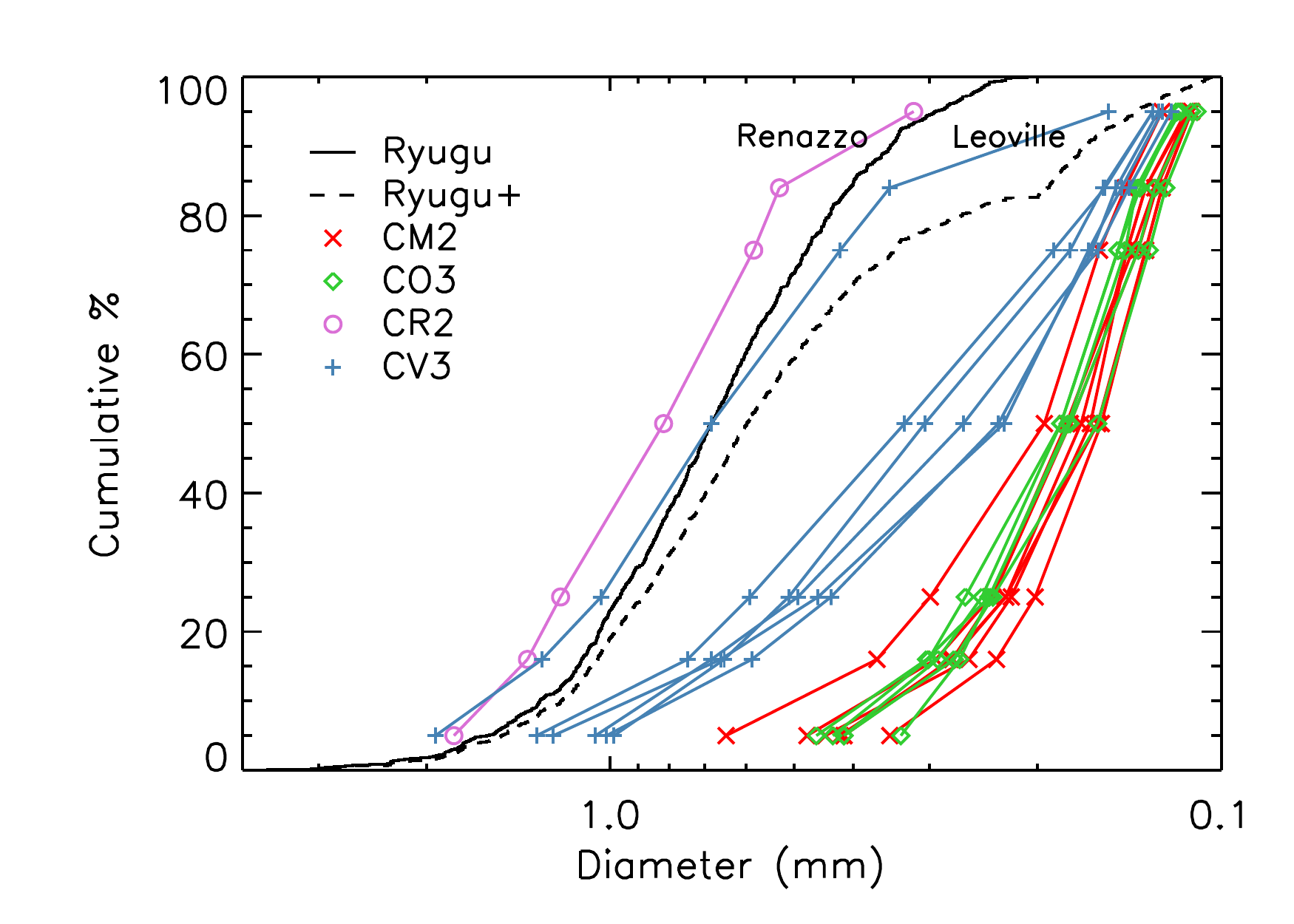}
	\caption{The cumulative size distribution of all inclusions mapped on the Ryugu rock at distance $< 25.5$~cm compared to those of meteorites in different carbonaceous chondrite groups \citep{KK78}. The dashed curve (``Ryugu+'') represents the Ryugu distribution with 200~additional inclusions with diameters randomly chosen between 0.1 and 0.2~mm. Two meteorite names are indicated.}
	\label{fig:size_distribution}
\end{figure}

The smallest and largest inclusions have sizes consistent with those of chondrules and refractory inclusions in CC meteorites, respectively. These two groups are distinguished not only by their size, but also on the basis of their morphology, with chondrules being spherical and refractory inclusions generally having an irregular shape. Unfortunately, the limited resolution of the MASCam images prevents us from verifying the spherical nature of chondrule candidates. Thus, given the absence of morphological and compositional information, we cannot clearly distinguish between chondrules and refractory inclusions. The inclusion size distribution may be diagnostic for CC group membership. The presence of abundant inclusions in the Ryugu rock excludes the CI group as an analog candidate, as members of this group typically lack inclusions. \citet{KK78} determined the size distribution of inclusions larger than 0.1~mm in 19~CC meteorites from 4~different groups (CM2, CO3, CR2, and CV3). The authors reported the inclusion maximum diameter as size parameter $\phi$, which relates to diameter as $d = 2^{-\phi}$~mm \citep{FW57}. We compare their measurements for all inclusions, regardless of shape or (suspected) type, to the Ryugu size distribution in Fig.~\ref{fig:size_distribution}. The CM2 and CO3 size distributions mostly overlap in a relatively narrow range, with the vast majority of their inclusions being smaller than 0.5~mm. CV3 inclusions are typically larger than CM2/CO3 inclusions, on average by a factor two. The inclusions of two meteorites are considerably larger than those of the others: Renazzo (CR2) and Leoville (CV3). The former was classified as CV2 by \citet{KK78}, but is now considered a CR2 \citep{C12a}. The Ryugu size distribution is close to these two meteorites, with a shape most similar to that of the Renazzo distribution. The \citet{KK78} distributions are supposed to be complete down to 0.1~mm diameter. But we likely underestimate the number of the smallest inclusions in the Ryugu rock (see Sec.~\ref{sec:size}). We therefore modified the Ryugu size distribution by adding 200~inclusions with a size between 0.1 and 0.2~mm, bringing the smallest size bin in the histogram in Fig.~\ref{fig:inclusion_area}b up to the level of its neighbor. Doing so changes the distribution curve only little, merely bringing its shape in better agreement with that of Leoville. It appears that the similarity of the Ryugu size distribution to the Renazzo and Leoville distributions is robust, which implies that the rock's inclusions are rather large for a CC. We do not suggest that these two meteorites are good analogs for the Ryugu rock, as their density is too high \citep{C97}.

Another diagnostic quantity is the inclusion abundance. We estimated that the areal abundance of small spherical inclusions is approximately equal to their volume abundance. We derived an inclusion abundance of around 7\%, which means a matrix abundance of 93\%. \citet{KK78} also estimated the abundance of the matrix, defined as including particles smaller than 0.1~mm. Table~\ref{tab:matrix_abundance} compares the Ryugu matrix abundance with that of the CC groups. The Ryugu abundance of 93\% is an upper limit, and agrees, in principle, with that of any group. However, the abundance is not likely to be underestimated by much, as the greatest uncertainty is associated with the number of smallest inclusions, which contribute only little area. The Ryugu matrix abundance therefore appears most similar to that of the CM2 group. The matrix abundance for the Renazzo and Leoville meteorites is 48\% and 65\%, respectively, considerably lower than that of the Ryugu rock. That said, it is not clear to what extent the size measurements of inclusions in MASCam images can be directly compared with the measurements on slabs of meteorite material by \citet{KK78}\footnote{To address the deficiencies in our analysis, we performed an experiment in which we imaged a number of CCs with a MASCam model. We will analyze these images in a similar way as the Ryugu rock in this paper, so the results may be directly compared. As our sample includes many meteorites that were also analyzed by \citet{KK78}, we will be able to verify that our methods yield comparable size distributions. The results of this experiment will be reported separately.}. We tentatively conclude that Ryugu inclusions are comparatively large, more similar to those associated with the CR2 and CV3 groups than the CM2 and CO3 groups. On the other hand, the matrix abundance is comparatively high, most similar to that seen in the CM2 group.

\begin{figure}
	\centering
	\includegraphics[width=11cm,angle=0]{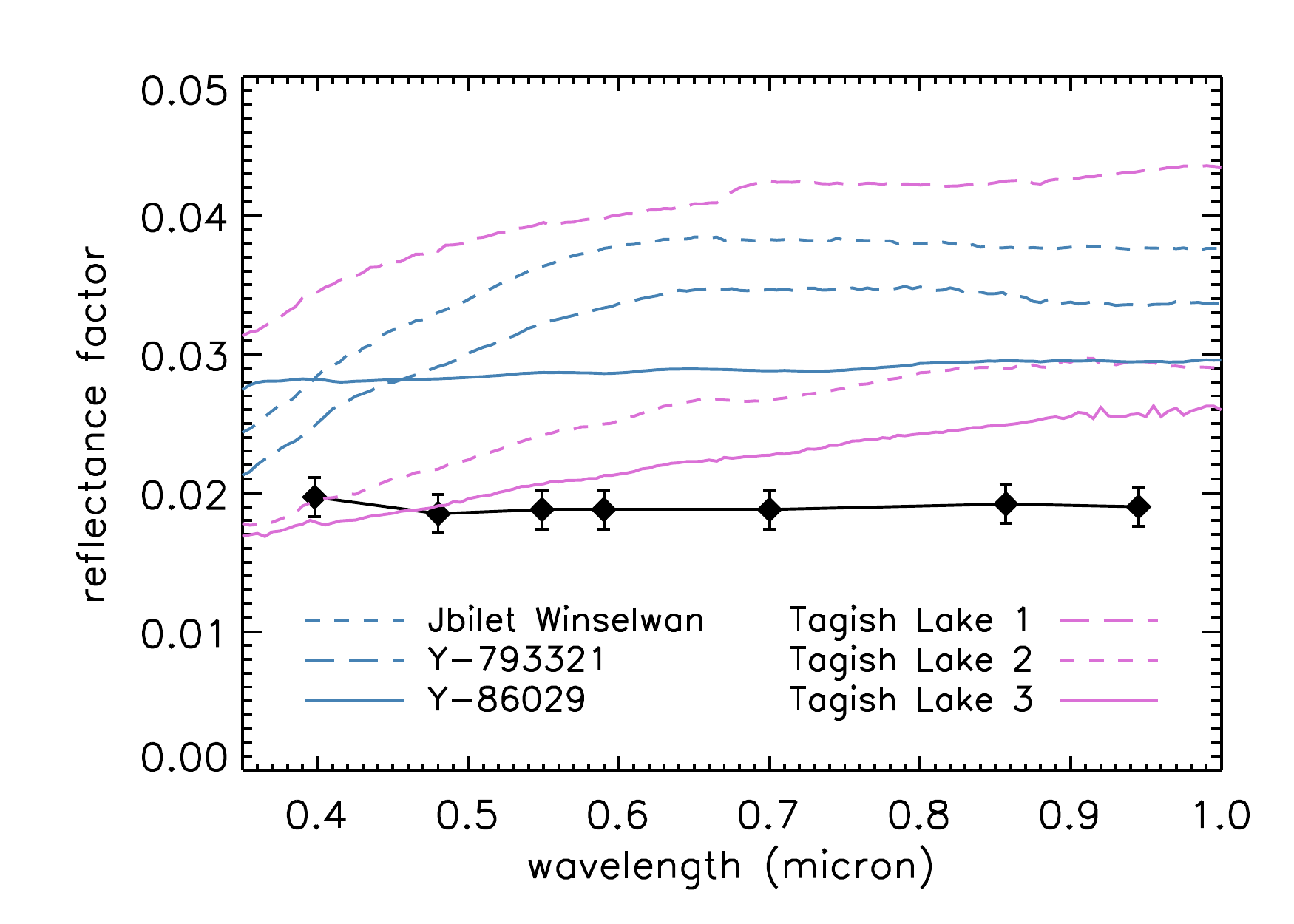}
	\caption{Reflectance spectra of three {\sc atcc}s and three Tagish Lake powder samples compared to that of Ryugu at the standard geometry of $(\iota, \epsilon, \alpha) = (30^\circ, 0^\circ, 30^\circ)$. The {\sc atcc} spectra (blue) are those in Fig.~3 of \citet{Su19}. The Tagish Lake~3 spectrum is that in Fig. 1 of the supplementary material of \citet{H01}. The Ryugu spectrum (black diamonds) is that in Fig.~20 in \citet{T20}. Details of the meteorite spectra are given in Table~\ref{tab:RELAB}.}
	\label{fig:meteorite_spectra}
\end{figure}

Lacking compositional information, the inclusion size distribution is not a sufficient criterion for group membership, and the implications of the observed inclusion color variations are unclear. Nevertheless, it appears that the Ryugu rock does not easily fit into any of the CC groups. Aqueously altered and thermally metamorphosed meteorites ({\sc atcc}) are plausible analogs because of their low reflectance \citep{C12c,Su19,K19}. Unfortunately, the size distribution and spectrophotometric properties of their inclusions have not been systematically documented. The Tagish Lake meteorite is also considered a possible analog \citep{J19}. Attractive properties are its low density and abundant inclusions \citep{H01,Z02}. But is it also a good spectral analog? If we adopt the criterion of \citet{Su19}, a similarly low reflectance over the visible wavelength range, it is. In Fig.~\ref{fig:meteorite_spectra} we compare the reflectance spectrum of three samples of Tagish Lake from the {\sc relab} database\footnote{\url{http://www.planetary.brown.edu/relabdocs/relab_disclaimer.htm}} (Table~\ref{tab:RELAB}) with that of Ryugu and the three {\sc atcc}s shown in Fig.~3B of \citet{Su19}. Tagish Lake spectrum~\#3 \citep{H01} is closer to the Ryugu spectrum than that of the {\sc atcc}s. The other two Tagish Lake spectra span the {\sc atcc} range. But spectrum~\#1 may have the highest overall reflectance because its sample was a pressed pellet. Pressing the surface is expected to increase the reflectance at the standard viewing geometry (e.g.\ \citealt{S14}). Spectral ambiguity also exists for the {\sc atcc}s: All spectra in Fig.~\ref{fig:meteorite_spectra} are for particular samples. The particulate spectrum of Y-86029 is flat, but a fragment of the same meteorite has a red spectral slope and is 30-40\% more reflective \citep{T20}. Meteorite fragments may be more representative for the Ryugu rock than powders, but a spectrum for a Tagish Lake fragment is not available in {\sc relab}. A detailed assessment of the 2.7~µm band is beyond the scope of this paper, but we do note that, for one Tagish Lake sample in {\sc relab}\footnote{Sample MT-TXH-025-L0, spectrum BKR1MT025L0. We note that this sample was pressed, which condition may have affected the reflective properties of the particulate.}, this band has a similar depth and shape as that of the heated Ivuna sample in \citet{H96}, which was adopted as the best match for Ryugu by \citet{K19}.

Meteorites with a reflectance as low as that of Ryugu are not common. Tagish Lake is unique and dark {\sc atcc}s appear to be relatively rare among the CCs. \citet{N05} identified 21 {\sc atcc}s among the CCs found on Antarctica prior to 1992, to which \citet{C12c} added 6 more. The Meteoritical Bulletin Database\footnote{\url{https://www.lpi.usra.edu/meteor/metbull.php}} lists 346~Antarctic CCs found prior to 1992. These numbers imply an abundance of  8\% for {\sc atcc}s in the pre-1992 Antarctic CC population. We can also approach the question of abundance in terms of meteorite mass. The 345~Antarctic CCs (one does not have mass listed in the database) have a total mass of 49.56~kg, half of which derives from a single meteorite (Y-791717). The total mass of the 27~{\sc atcc}s is 4.90~kg, which represents 10\% of the total mass. While ours is not a comprehensive assessment, these abundances suggest that {\sc atcc}s are relatively rare, with the caveats that many {\sc atcc}s may not yet have been recognized as such and that not all are as dark as Ryugu \citep{C12c}. On the other hand, C-type asteroids as dark as Ryugu are common \citep{BS00,S10}. In fact, all three spacecraft encounters with C-type asteroids have yielded similarly low geometric albedos in the visible: $0.047 \pm 0.005$ for 253~Mathilde \citep{C99}, $0.045 \pm 0.002$ for Ryugu \citep{Su19}, and $0.044 \pm 0.002$ for 101955~Bennu \citep{L19}. If {\sc atcc}s are a good analog for low-albedo C-type asteroid material, we would expect them to be common. They may exist in abundance in space but not make it to Earth's surface. Then, the apparent rarity of {\sc atcc}s and Tagish Lake's unique status are consistent with the idea that Ryugu-type meteorites are too fragile to survive atmospheric entry \citep{G19}.

\section*{Acknowledgments}

This study was supported by JSPS International Planetary Network. This research utilizes spectra acquired by Takahiro Hiroi, Michael Zolensky, Katsuhito Ohtsuka, Carle Pieters, and David Kring with the {\sc nasa relab} facility at Brown University. We thank two anonymous reviewers for their helpful comments. The recalibrated MASCam images and associated calibration frames are available at \url{http://europlanet.dlr.de/Hayabusa2/MASCOT/}.

\newpage

\bibliography{Ryugu}{}
\bibliographystyle{aasjournal}


\newpage
\clearpage

\begin{table}
	\centering
	\caption{Image sets acquired during the two nights on Ryugu.}
	\label{tab:timeline}
	\begin{tabular}{lllll}
		\hline
		\hline
		Night & Set & Location & ID & $T$ [$^\circ$C] \\
		\hline
		1 & 5 & MP1a & 350-355 & $30$ \\
		1 & 6 & MP1a & 400-405 & $27$ \\
		1 & 7 & MP1a & 450-455 & $17$ \\
		1 & 8 & MP1a & 450-455 & $10$ \\
		1 & 9 & MP1b & 450-455 & $4$ \\
		2 & 11 & MP2b & 650-655 & $26$ \\
		2 & 12 & MP2c & 700-705 & $14$ \\
		2 & 13 & MP2c & 750-755 & $5$ \\
		2 & 14 & MP2c & 750-755 & $-4$ \\
		2 & 15 & MP2c & 750-755 & $-11$ \\
		\hline
	\end{tabular}
\tablecomments{ID refers to image number, $T$ is the detector temperature, and locations are from \citet{S19a}. The image ID is not unique for acquisition sequences that were repeated.}
\end{table}

\begin{table}
	\centering
	\caption{Details of the images in set~13, 14, and 15 in order of acquisition.}
	\label{tab:set14}
	\begin{tabular}{llr}
		\hline
		\hline
		ID & LED & $t_{\rm exp}$ [msec] \\
		\hline
		750 & -  & 962.1000 \\
		751 & red  & 2946.3778 \\
		752 & green  & 2946.3778 \\
		753 & blue  & 2946.3778 \\
		754 & infrared & 2946.3778 \\
		755 & -  & 0.2138 \\
		\hline
	\end{tabular}
\end{table}

\begin{table}
	\centering
	\caption{Revised responsivity factors $R_i$ for the LED colors.}
	\label{tab:responsivities}
	\begin{tabular}{llll}
		\hline
		\hline
		$i$ & LED & $\lambda_{\rm eff}$ & $R$ \\
		& & [nm] & [m$^2$~sr~mJ$^{-1}$] \\
		\hline
		1 & blue & $471^{+6}_{-16}$ & $109.7 \pm 1.2$ \\
		2 & green & $532^{+12}_{-24}$ & $130.7 \pm 1.3$ \\
		3 & red & $630^{+10}_{-7}$ & $127.3 \pm 1.4$ \\
		4 & infrared & $809^{+18}_{-17}$ & $95.4 \pm 1.3$ \\
		\hline
	\end{tabular}
\tablecomments{The range for the effective wavelength $\lambda_{\rm eff}$ is the {\sc fwhm}.}
\end{table}

\begin{table}
	\centering
	\caption{Matrix abundance in the Ryugu rock compared to estimates for different carbonaceous chondrite groups \citep{KK78}.}
	\label{tab:matrix_abundance}
	\begin{tabular}{llllll}
		\hline
		\hline
		& CM2 & CO3 & CR2 & CV3 & Ryugu \\
		\hline
		matrix area\% & $93 \pm 3 (7)$ & $74 \pm 9 (5)$ & $48 (1)$ & $58 \pm 8 (6)$ & $<93$ \\
		\hline
	\end{tabular}
\tablecomments{The number of meteorites on which the estimate is based is given in brackets.}
\end{table}

\begin{table}
	\centering
	\caption{Details for {\sc relab} meteorite spectra shown in Fig.~\ref{fig:meteorite_spectra}.}
	\label{tab:RELAB}
	\begin{tabular}{llll}
		\hline
		\hline
		Name in legend & Spectrum ID & Sample ID & Sample condition \\
		\hline
		Jbilet Winselwan & C1MT312A & MT-DAK-312-A & Particulate, ground, sorted \\
		Y-793321 & C1MP131 & MP-KHO-131 & Chip \\
		Y-86029 & C3MP111 & MP-TXH-111 & Particulate, ground, dry-sieved \\
		Tagish Lake 1 & C1MT25L2 & MT-TXH-025-L2 & Particulate, pressed \\
		Tagish Lake 2 & C1MT237C & MT-S1S-237-C & Particulate, ground, dry-sieved \\
		Tagish Lake 3 & C2MT11 & MT-MEZ-011 & Particulate, ground, dry-sieved \\
		\hline
	\end{tabular}
\end{table}

\newpage
\clearpage

\end{document}